\documentclass{aastex6}
\usepackage{mathrsfs}
\usepackage{amsmath}
\usepackage{color}
\usepackage{xcolor}

\begin{document}

\title{Relativistic Viscous Radiation Hydrodynamic Simulations of Geometrically Thin Disks: I. Thermal and Other Instabilities}
\shorttitle{Viscous Radiative Simulations of Thin Disks I.}
\shortauthors{Fragile et al.}
\author{P. Chris Fragile\footnotemark[1], Sarina M. Etheridge,}
\affil{Department of Physics and Astronomy, College of Charleston, Charleston, SC 29424, USA; fragilep@cofc.edu}
\author{Peter Anninos,}
\affil{Lawerence Livermore National Laboratory, P.O. Box 808, Livermore, CA 94550, USA}
\author{Bhupendra Mishra\footnotemark[1]}
\affil{JILA, University of Colorado and National Institute of Standards and Technology, 440 UCB, Boulder, CO 80309-0440, USA}
\author{W{\l}odek Klu\'zniak\footnotemark[1]}
\affil{Nicolaus Copernicus Astronomical Center, Bartycka 18, Warsaw, 00-716, Poland}

\footnotetext[1]{Kavli Institute for Theoretical Physics, Santa Barbara, CA.}

\begin{abstract}
We present results from two-dimensional, general relativistic, viscous, radiation hydrodynamic numerical simulations of Shakura-Sunyaev thin disks accreting onto stellar mass Schwarzschild black holes. We consider cases on both the gas- and radiation-pressure-dominated branches of the thermal equilibrium curve, with mass accretion rates spanning the range from $\dot{M} = 0.01 L_\mathrm{Edd}/c^2$ to $10 L_\mathrm{Edd}/c^2$. The simulations directly test the stability of this standard disk model on the different branches. We find clear evidence of thermal instability for all radiation-pressure-dominated disks, resulting universally in the vertical collapse of the disks, which in some cases then settle onto the stable, gas-pressure-dominated branch.  Although these results are consistent with decades-old theoretical predictions, they appear to be in conflict with available observational data from black hole X-ray binaries. We also find evidence for a radiation-pressure-driven instability that breaks the unstable disks up into alternating rings of high and low surface density on a timescale comparable to the thermal collapse. Since radiation is included self-consistently in the simulations, we are able to calculate lightcurves and power density spectra (PDS). For the most part, we measure radiative efficiencies (ratio of luminosity to mass accretion rate) close to 6\%, as expected for a non-rotating black hole. The PDS appear as broken power laws, with a break typically around 100 Hz. There is no evidence of significant excess power at any frequencies, i.e. no quasi-periodic oscillations are observed. 
\end{abstract}

\maketitle

\section{Introduction}

From the earliest work on thin ($H/R \ll 1$) accretion disks based on the $\alpha$-viscosity prescription \citep{Shakura73}, there have been notable problems whenever the steady-state solution calls for the disk to be radiation pressure dominated, as it will be above a certain mass accretion rate, $\dot{m} = \dot{M}/\dot{M}_\mathrm{Edd} \gtrsim 0.02$, where $\dot{M}_\mathrm{Edd} = L_\mathrm{Edd}/c^2 = 1.3 \times 10^{17} M/M_\odot$ g s$^{-1}$ is the Eddington mass accretion rate, and inside a certain radius, $r \lesssim 100 \dot{m}^{16/21} r_g$, where $r_g = GM/c^2$.  Specifically, the solution is predicted to be both thermally \citep{Shakura76} and viscously \citep{Lightman74} unstable.  The thermal instability arises owing to different dependencies of the disk heating, $Q^+$, and cooling, $Q^-$, on mid-plane temperature, $T_c$, such that small deviations in the temperature are expected to lead to runaway heating or cooling.  Similarly, the viscous instability arises from an inverse dependence of the vertically integrated stress, $W_{R\phi}$, on surface density, $\Sigma$, leading to an anti-diffusion equation for $\Sigma$ such that the disk breaks up into alternating rings of high $\Sigma$, low $W_{R\phi}$, and low $\Sigma$, high $W_{R\phi}$.

This is particularly puzzling in light of observations of black hole X-ray binaries (BHXRBs). In these systems, the spectra look most disk-like (soft, with a prominent thermal bump around 1 keV) and stable (rms variability $\lesssim 3$\%) whenever $L = 0.1-0.2 L_\mathrm{Edd}$ \citep[e.g.][]{vanderKlis04,Done07}, corresponding to $\dot{m} = 2-3$ for a standard efficiency of $\eta = 0.057$, precisely when such disks are predicted to be {\it unstable}. So, the first puzzle one would like to solve is precisely when are these disks unstable. 

The second point of interest is to find out what the non-linear manifestations of these instabilities are in real accretion disks and their consequences over periods comparable to a viscous timescale.  One way to get a handle on these questions is through direct numerical simulations.  At a minimum, this requires evolution of the equations of viscous, {\em radiation} hydrodynamics.  Alternatively, since the true source of the ``viscosity'' in many accreting systems is thought to be turbulence driven by the magneto-rotational instability \citep[MRI][]{Balbus92,Balbus98}, one could perform radiation {\em magnetohydrodynamic} (MHD) simulations.  The codes required to perform such direct numerical simulations have only recently become available \citep[e.g.][]{Ohsuga05,Krolik07,Jiang13,Sadowski13,McKinney14}, and we apply one such code, {\em Cosmos++} \citep{Anninos05,Fragile14}, in this work.

The thermal stability of radiation-pressure-dominated accretion disks were first numerically studied in the context of limit-cycle-behavior using time-dependent, one-dimensional, height-integrated disk structure equations \citep{Honma91,Honma92,Janiuk02}. Local, shearing box simulations that treat a small patch of the accretion disk, but include vertical and azimuthal structure, followed \citep{Agol01,Turner02}. Although there was an early claim of thermal stability \citep{Hirose09}, the consensus now is that, indeed, radiation-pressure-dominated disks are thermally unstable \citep{Jiang13,Ross17}. We confirmed this behavior, and in addition found evidence for the viscous instability, in our own global radiation MHD simulations \citep{Mishra16}. Global simulations are critical to capturing the viscous instability, as it requires a larger radial range than is typically treated in shearing box simulations. They also offer a better potential for comparison with observations. However, the disks simulated in the \citet{Mishra16} paper were not true Shakura-Sunyaev disks and did not begin on the thermal equilibrium curve.

In the present work, we set up numerical experiments to directly test the thermal and viscous stability of Shakura-Sunyaev disks around Schwarzschild black holes.  We do this by adding a full, relativistic Navier-Stokes treatment to the momentum and energy evolution equations in our {\em Cosmos++} computational astrophysics code.  This allows us to evolve the disk with the exact $\alpha$-viscosity treatment of \citet{Shakura73}.  As expected the radiation-pressure-dominated cases that we consider show unmistakable evidence for thermal instability, plus another instability, that is possibly radiation-driven, but likely not the viscous instability. Because we include radiation in the simulations, we are able to directly measure the luminosities and efficiencies associated with our simulated disks, as well as characterize their variability. Each of these results are described, in turn, in Sections \ref{sec:thermal} -- \ref{sec:luminosity}. Before getting to those results, though, we describe our numerical methods in Section \ref{sec:methods}.  Finally, in Section \ref{sec:conclusion}, we present some discussions and conclusions. Unless otherwise noted, standard index notation is used for labeling spacetime coordinates: repeated indices represent summations, with Latin and Greek indices running over the spatial and 4-space dimensions, respectively.

\section{Numerical Methods}
\label{sec:methods}

All of the simulations presented in this paper use the {\em Cosmos++} computational astrophysics code \citep{Anninos05,Fragile12,Fragile14} to numerically evolve the equations of general relativistic radiative, viscous hydrodynamics, as described in the next section.  {\em Cosmos++} is a parallel, multi-dimensional, fully covariant, modern object-oriented (C++) radiation hydrodynamics and magnetohydrodynamics (MHD) code, written to support structured and unstructured adaptively-refined meshes, and for both Newtonian and general relativistic astrophysical applications. For this work, we utilize the High Resolution Shock Capturing (HRSC) scheme as described in \citet{Fragile12}, which shares many elements with the non-oscillatory central difference (NOCD) method presented in \citet{Anninos03}. In the next sections, we describe how we treat the viscosity and radiation in these simulations.

\subsection{Relativistic, Viscous, Radiation Hydrodynamics}

The contravariant stress energy tensor for a viscous fluid can be represented as a linear combination of the hydrodynamic and viscous contributions as
\begin{equation}
T^{\alpha\beta} =
         \underbrace{\rho h u^\alpha u^\beta + P_\mathrm{gas} g^{\alpha\beta}}_\mathrm{hydro}
           - \underbrace{(Q_B/c^2) u^\alpha u^\beta - Q_B g^{\alpha\beta} - Q^{\alpha\beta}_S}_\mathrm{viscous} ~.
\end{equation}
Collecting terms, this can be written more conveniently as
\begin{equation}
T^{\alpha\beta} =
         (\rho h - Q_B/c^2) u^\alpha u^\beta 
         + (P_\mathrm{gas}-Q_B) g^{\alpha\beta} 
         - Q^{\alpha\beta}_S ~,
\label{eqn:tmn}
\end{equation}
where $\rho$ is the fluid rest-mass density, $h=1+\epsilon/c^2 + P_\mathrm{gas}/(\rho c^2)$ is the specific enthalpy, $c$ is the speed of light, $u^\alpha$ is the contravariant four-velocity, $P_\mathrm{gas}$ is the fluid pressure [for an ideal gas $P_\mathrm{gas}=(\Gamma-1)e$, where $e=\rho\epsilon$ is the fluid internal energy density and $\Gamma$ is the adiabatic index], $Q_B$ is the bulk (scalar) viscosity, $Q^{\alpha\beta}_S$ is the (symmetric) shear tensor viscosity, and $g_{\alpha\beta}$ is the curvature metric.  The viscous forces are explicitly split into bulk and shear viscosities for implementation convenience. We emphasize the $Q_B$ and $Q^{\alpha\beta}_S$ may include both molecular and artificial (for shock capturing) viscosity contributions.

For the radiation, we employ a covariant formulation of the $\bf{M}_1$ closure scheme \citep{Levermore84,Sadowski13}, which assumes that the radiation is isotropic in the radiation rest frame.  If $E_R$ and $u_R^i$ represent the radiation energy density in the radiation rest frame and the spatial components of the radiation rest frame 4-velocity, respectively, then the radiation stress tensor becomes \citep{Sadowski13,Fragile14}
\begin{equation}
R^{\alpha \beta} = \frac{4}{3} E_R u^{\alpha}_R u^{\beta}_R + \frac{1}{3} E_R g^{\alpha \beta}~.
\label{eq:radstress}
\end{equation}
The $\bf{M}_1$ closure scheme has been shown to underestimate the flux coming out of plane-parallel atmospheres compared to semi-analytic solutions (Yan-fei Jiang, private communication). For our work, this means that we are most likely underestimating the radiative cooling of our disks. Since we find that our radiation-pressure-dominated disks are thermally unstable, usually with cooling dominating over heating, this suggests that our conclusions would be even stronger if we were using a more accurate radiation hydrodynamics scheme.

The eight equations of radiation hydrodynamics (energy plus three components of momentum for both the fluid and radiation fields) are derived from the conservation of fluid and radiation stress energy: $\nabla_\mu T^\mu_{\ \nu} = \partial_\mu T^\mu_{\ \nu} +  \Gamma^\mu_{\alpha\mu} T^\alpha_{\ \nu} - \Gamma^\alpha_{\mu\nu} T^\mu_{\ \alpha} = G_\nu$ and $\nabla_\mu R^\mu_{\ \nu} = -G_\nu$, where $G_\alpha$ is the radiation 4-force density, which couples the fluid and radiation fields \citep{Mihalas84}. In this work, we generalize the radiation four-force density from that presented in \citet{Fragile14} to include a Compton scattering component.  In this work
\begin{equation}
G^\mu = -\rho \left(\kappa_\mathrm{F}^\mathrm{a} + \kappa^\mathrm{s}\right) R^{\mu \nu}u_{\nu} - \rho\left\{\left[ \kappa^\mathrm{s} + 4 \kappa^\mathrm{s}\left(\frac{T_\mathrm{gas} - T_\mathrm{rad}}{m_e}\right) + \kappa_\mathrm{F}^\mathrm{a} - \kappa_\mathrm{J}^\mathrm{a} \right] R^{\alpha \beta} u_{\alpha} u_{\beta} + \kappa_\mathrm{P}^\mathrm{a} a_R T_\mathrm{gas}^4 \right\} u^{\mu}~,
\label{eq:Gmu}
\end{equation}
and we assume Kramers-type opacity laws. Since free-free absorption is the most relevant atomic absorption process for stellar mass black hole accretion disks, the appropriate Planck and Rosseland means (for solar metallicity and a hydrogen mass fraction of $X=0.7$) are $\kappa_\mathrm{P}^\mathrm{a} = 6.4\times 10^{22} T^{-7/2}_\mathrm{K} \rho_\mathrm{cgs} ~\mathrm{cm}^2~\mathrm{g}^{-1}$ and $\kappa_\mathrm{R}^\mathrm{a} = 1.6\times 10^{21} T^{-7/2}_\mathrm{K} \rho_\mathrm{cgs} ~\mathrm{cm}^2~\mathrm{g}^{-1}$ \citep[e.g.][]{Hirose09}, respectively, where $T_\mathrm{K}$ is the ideal gas temperature of the fluid in Kelvin and $\rho_\mathrm{cgs}$ is the density in g cm$^{-3}$.  In this work, we assume the flux mean, $\kappa_\mathrm{F}^\mathrm{a}$, is the same as the Rosseland mean and the J-mean, $\kappa_\mathrm{J}^\mathrm{a}$, is the same as the Planck mean. Because we neglect ionization/recombination processes and composition effects, we use a constant electron scattering opacity, $\kappa^\mathrm{s} = 0.2(1+X) = 0.34 ~\mathrm{cm}^2~\mathrm{g}^{-1}$.  In this work, we assume that the electron-ion equilibration time is sufficiently short for the electrons to be at the same temperature as the ions \citep[see][for discussions of the validity and impact of this choice]{Ressler17,Sadowski17}. In addition to the conservation of energy and momentum, we also require an equation for the conservation of mass: $\nabla_\mu(\rho u^\mu) = \partial_t(\sqrt{-g} u^0\rho) + \partial_i (\sqrt{-g} u^0 \rho V^i) = 0$.  Note that we do not presently include an equation for the conservation of photon number, though this choice is not expected to significantly affect our results \citep{Sadowski15}.

\subsubsection{Conservation Equations}

The conservation of stress energy for the fluid can be expanded and rearranged as
\begin{equation}
\partial_t(\sqrt{-g} T^0_\nu) + \partial_i(\sqrt{-g} T^i_\nu)
    = \sqrt{-g} T^\mu_\sigma \ \Gamma^\sigma_{\mu\nu} + \sqrt{-g} G_\nu ~.
\end{equation}
By defining the total energy density and momentum density from the full stress tensor as ${\cal E} = - \sqrt{-g} {T}^0_0$ and ${\cal S}_j = \sqrt{-g} {T}^0_j$, the corresponding conservation equations in this framework are
\begin{equation}
\partial_t {\cal E} + \partial_i( - \sqrt{-g} T^i_0)
    = - \sqrt{-g} T^\mu_\sigma \ \Gamma^\sigma_{\mu 0}  - \sqrt{-g}~G_0 
\end{equation}
or
\begin{equation}
\partial_t {\cal E} + \partial_i({\cal E} V^i)
    + \partial_i [ \sqrt{-g} (P - Q_B - Q^0_0) V^i
                  +\sqrt{-g} \ Q^i_0 ]
    = - \sqrt{-g} T^\mu_\sigma \ \Gamma^\sigma_{\mu 0} - \sqrt{-g}~G_0 ~,
\end{equation}
for energy, and
\begin{equation}
\partial_t {\cal S}_j + \partial_i( \sqrt{-g} T^i_j)
    = \sqrt{-g} T^\mu_\sigma \ \Gamma^\sigma_{\mu j} + \sqrt{-g}~G_j 
\end{equation}
or
\begin{equation}
\partial_t {\cal S}_j + \partial_i({\cal S}_j V^i)
    + \partial_i \{ \sqrt{-g} [(P - Q_B) g^0_j - Q^0_j] V^i
                  +\sqrt{-g} \ Q^i_j \}
    = \sqrt{-g} T^\mu_\sigma \ \Gamma^\sigma_{\mu j} + \sqrt{-g}~G_j ~,
\end{equation}
for momentum.  Likewise, defining the radiation total energy density and radiation momentum density as ${\cal R} = \sqrt{-g}R^0_0$ and ${\cal R}_j = \sqrt{-g} R^0_j$, the corresponding conservation equations are 
\begin{equation}
 \partial_t {\cal R} + \partial_i \left(\sqrt{-g}~R^i_0\right) =
      \sqrt{-g}~R^\alpha_\beta~\Gamma^\beta_{0 \alpha} - \sqrt{-g}~G_0 ~,
    \label{eqn:rad_en}
\end{equation}
and
\begin{equation}
 \partial_t {\cal R}_j + \partial_i \left(\sqrt{-g}~R^i_j\right) =
      \sqrt{-g}~R^\alpha_\beta~\Gamma^\beta_{j \alpha} - \sqrt{-g}~G_j ~,
    \label{eqn:rad_mom} 
\end{equation}
respectively. These equations are discretized using a second-order finite volume representation, with third-order piecewise-parabolic interpolations for the flux reconstructions, and evolved forward using a second-order, low-storage Euler time-stepping scheme.

\subsubsection{Primitives}

At the end of each time cycle a series of coupled nonlinear equations are solved to extract the primitive fields from the evolved conserved fields, using the semi-implicit method described in \citet{Fragile14}.  Essentially the method utilizes Newton-Raphson iteration to solve a linear matrix system constructed from the primitive field dependency of all of the conserved quantities. Thus within each iteration one constructs a ($9\times9$ for the case of radiation hydrodynamics) Jacobian matrix $A_{ij} = \partial U^i/\partial P^j$ evaluated using initial guesses for the primitive solutions. Here $U^i \equiv \{D, {\cal E}, {\cal S}_k, {\cal R}, {\cal R}_k\} = \sqrt{-g} \{u^0\rho, - T^0_0, T^0_k, R^0_0, R^0_k\}$ is a vector list of conserved fields, while $P^j \equiv \{\rho, \epsilon, \widetilde{u}^k, E_R, \widetilde{u}^k_R\}$ is a vector list of corresponding primitive fields. We introduce $\widetilde{u}^k = u^k - u^0 g^{0k}/g^{00}$, with $u^0 = \sqrt{-g^{tt}}\gamma$, as the primitive velocity in place of $v^2$, where $\gamma = \sqrt{1+g_{ij} \widetilde{u}^i \widetilde{u}^j}$. The energy and momentum matrix elements are of particular interest since they are based on the stress energy tensor, ${\cal E}=-\sqrt{-g} T^0_0$, ${\cal S}_k = \sqrt{-g} T^0_k$, and differ from ideal hydrodynamics by the addition of viscous terms. Keeping in spirit with the quasi-static approach to solving the primitive value problem, the stress tensor built from primitive fields within each Newton iteration includes viscous terms, but the row-wise Jacobian matrix elements neglect their variation as a higher order correction. Viscous terms are thus assumed constant during the primitive solve, but are updated during each hydrodynamic cycle with the same high order temporal and spatial discretization as the radiation hydrodynamics. The one exception is the treatment of the 4-velocity time derivatives, which are approximated as first order differences over each sub-cycle step.

\subsubsection{Viscosity Stress Tensor}
Physical (molecular) viscosity takes the generic form
\begin{equation}
Q^{\alpha\beta} = \mu h^{\sigma (\beta} \nabla_{\sigma} u^{\alpha )}
               + \left(\mu_B - \frac{\mu}{3}\right) 
                 h^{\alpha\beta} \nabla_\sigma u^\sigma ~,
\label{eqn:Qab}
\end{equation}
where $\mu$ and $\mu_B$ are the shear and bulk coefficients, respectively, in units of mass/(length$\times$time), and $h^{\alpha\beta} = g^{\alpha\beta} + u^\alpha u^\beta/c^2$ is the projection operator. The symmetric tensor $Q^{\alpha\beta}$ satisfies the orthogonality condition $u^\mu Q_{\mu\nu} = 0$, offering a convenient way to compute its time-time and time-space components.

Following the standard theory of thin disks, the shear viscosity coefficient is calculated as 
\begin{equation}
\mu=\nu\rho=\alpha\rho c_s H ~,
\end{equation}
where $c_s$ is the thermal sound speed (including both gas and radiation contributions), $H$ is the disk height, and $\alpha$ is the Shakura-Sunyaev viscosity parameter.  In this work, $\alpha$ is assumed to be a constant, while $c_s$ and $H$ are calculated from the local conditions.  Specifically, $c_s = \sqrt{P_\mathrm{tot}/\rho}$, where $P_\mathrm{tot} = P_\mathrm{gas} + P_\mathrm{rad}$ is the total pressure, and  $H = c_s/V^\phi$, although we also tested one case where $H = c_s/\Omega$, where $\Omega$ is the Keplerian angular frequency. Note that this height, $H$, is simply used to scale $\mu$; it does not restrict where viscosity is applied. In all cases, the effective height is limited to $H < 0.1 r$; this was done to prevent having very large viscosity in the background gas where $V^\phi$ fluctuates considerably. We also include a bulk viscosity of the same form, $\mu_\beta = \nu\rho$, whose main effect is to help damp out spurious sound waves that come from mismatches in the initial conditions and outer boundary.

For added numerical stability and to preserve causality, we restrict the hydrodynamical timestep to be less than the minimum calculated from the local viscosity speed ($\Delta t < \rho \Delta x^2/\mu$).  We further limit the viscous source tensor to
\begin{equation}
Q^\mu_\nu = \frac{Q^\mu_\nu}{\mathrm{max}\left[1,\sqrt{Q^\alpha_\beta Q^\beta_\alpha}/(\chi_Q P_\mathrm{tot}) \right]} ~,
\end{equation}
where $\chi_Q = 1.2$ is a parameter of order unity that controls the amount of limiting. This is useful for preventing the anisotropic component of viscosity from exceeding the isotropic pressure.

\subsection{Simulation Setup}

To initialize our simulations, we start from the \citet{Novikov73} generalization of the Shakura-Sunyaev thin disk.  As we are only considering a limited radial range, we do not require all three regions of the solution.  Instead, whenever we want to initialize a gas-pressure-dominated case (appropriate for $\dot{m} \lesssim 0.02$), we only initialize the so-called ``middle'' region (for this $\dot{m}$, the so-called ``inner'' region only exists inside the innermost stable circular orbit, or ISCO).  On the other hand, whenever we want to initialize a radiation-pressure-dominated case, we only initialize the ``inner'' region.  We follow the form of the Novikov-Thorne solutions given in \citet{Abramowicz13}, although all we actually require are the radial dependencies of $H(R)$ and $\rho_0(R)$, where $\rho_0$ is the mid-plane density.  We also include the small radial velocity, $V^R(R)$, from \citet{Penna12}, associated with the slow radial drift of material through the disk.

For the vertical profile, we solve for the vertical hydrostatic equilibrium, either assuming an isothermal disk in the gas-pressure-dominated case or a polytropic EOS in the radiation-pressure-dominated case. For the former, the solution yields
\begin{equation}
\rho(R,z) = \rho_0 \mathrm{e}^{-z^2/2H^2} 
\end{equation}
and
\begin{equation}
P_\mathrm{tot}(R,z) = \frac{G M_\mathrm{BH} H^2}{R^3}\rho(R,z)~.
\end{equation}
In the latter case,
\begin{equation}
\rho(R,z) = \rho_0 \left[1 - \frac{z^2}{2H^2}\right]^{1/(\Gamma_\mathrm{NT}-1)}
\end{equation}
and
\begin{equation}
P_\mathrm{tot}(R,z) = \kappa \rho^{\Gamma_\mathrm{NT}} ~,
\end{equation}
where
\begin{equation}
\kappa = \frac{GM_\mathrm{BH} H^2}{\Gamma_\mathrm{NT} (\Gamma_\mathrm{NT}-1) \rho_0^{\Gamma_\mathrm{NT}-1} R^3} ~.
\end{equation}
For the radiation-pressure-dominated cases, $\Gamma_\mathrm{NT} = 4/3$. 

Assuming the gas and radiation are in local thermodynamic equilibrium for the initial, analytic solution, we partition the pressure as
\begin{equation}
P_\mathrm{tot} = P_\mathrm{gas} + P_\mathrm{rad} = \frac{k_\mathrm{b}\rho T_\mathrm{gas}}{\bar{m}} + \frac{1}{3}a_\mathrm{R} T_\mathrm{gas}^4 ~,
\end{equation}
where $\bar{m} = 0.615 m_H$ and $a_\mathrm{R} = 4\sigma/c$ is the radiation constant.  We can now solve for $T_\mathrm{gas}(R,z)$.  The initial azimuthal velocity is taken to be Keplerian, $V^\phi(R) = \Omega$.  Note that we neglect the additional corrections to the Novikov-Thorne solution suggested by \citet{Penna12}, but since we are just using these conditions to initialize our simulations, this should not matter too much.  For the background, we initialize a cold ($e = 10^{-12} \rho_\mathrm{max} r^{-5/6}$), low density ($\rho = 10^{-10} \rho_\mathrm{max} r^{-5/2}$), static (${V}^k = 0$) fluid. 

The temperature found above is also used to set the radiation field. In the frame of the fluid, the radiation energy density is
\begin{equation}
E_\mathrm{rad} = a_R T_\mathrm{gas}^4 ~,
\end{equation}
while the flux, $F^i$, is initially set equal to the gradient of this quantity.  To get the radiation density in the radiation rest frame, $E_R$, and the radiation rest-frame four-velocity, $u_R^\mu$, we follow the transformation procedure outlined in \citet{Sadowski13}.

For simulation S01E, the disk initially extends from $r = 6 r_\mathrm{g}$ to the outer domain boundary, $r_\mathrm{max}$. For all other simulations, the disk is initially truncated at $r = 7.5 r_\mathrm{g}$. We chose to do this because the analytic solution becomes extremely thin close to the ISCO and leads to very large gradients that cause numerical difficulties at the start of the simulations. Instead, we let these simulations fill in this region as they evolve. For the inner boundary, we do not go all the way to the black hole event horizon, $r_\mathrm{BH}$, but rather truncate our domain at some intermediate radius, $r_\mathrm{min}$ (between $r_\mathrm{BH}$ and $r_\mathrm{ISCO}$).  This is done strictly to reduce computational cost. We utilize logarithmic spacing of the form
\begin{equation}
x_1 = 1 + \ln \left(\frac{r}{r_\mathrm{BH}}\right)
\end{equation}
in the radial direction.  We also only consider a small wedge in $\theta$, with resolution concentrated toward the mid-plane by a function of the form
\begin{equation}
\theta = x_2 + \frac{1}{2} [1 - q] \sin(2 x_2) ~,
\end{equation}
with values of $q$ provided in Table \ref{tab:params}.  All simulations that we report in this paper use a resolution of $256 \times 192$. Although this may seem like a modest resolution, the combination of a limited range in $\theta$ and the concentrated latitude coordinate results in an effective resolution in the vertical direction as small as 24 m in some simulations. Lower resolution test simulations exhibit the same basic behaviors, suggesting that our conclusions are robust. The inner, top, and bottom boundaries use outflow conditions, where matter and radiation are only allowed to flow off the grid.  At the outer boundary we hold the ghost zone values at those prescribed by the Novikov-Thorne solution, though material (and radiation) can still leave the grid through this boundary.

\begin{deluxetable}{cccccccc}
\tablecaption{Simulation Models and Parameters \label{tab:params}}
\tablecolumns{7}
\tablehead{
\colhead{Name} & \colhead{$\dot{m}$} & \colhead{$r_\mathrm{min}$} & \colhead{$r_\mathrm{max}$} & \colhead{$\theta_\mathrm{max} - \theta_\mathrm{min}$} & \colhead{$q$} & \colhead{$t_\mathrm{stop}$}  \\
 &  & \colhead{($GM/c^2$)} & \colhead{($GM/c^2$)} & \colhead{(rad)} &  & \colhead{($GM/c^3$)} }
\startdata
S01E & 0.01 & 5 & 20 & 0.289 & 0.1 & 80,485 \\
S1E & 1 & 5 & 20 & 0.401 & 0.1 &  80,652\\
S3E & 3 & 4 & 40 & 0.476 & 0.3 & 42,964 \\
S3Ep & 3 & 4 & 40 & 0.476 & 0.3 & 24,921 \\
S10E & 10 & 4 & 40 & 0.871 & 0.3 & 27,492 
\enddata
\end{deluxetable}

\begin{figure}
\includegraphics[width=0.48\columnwidth]{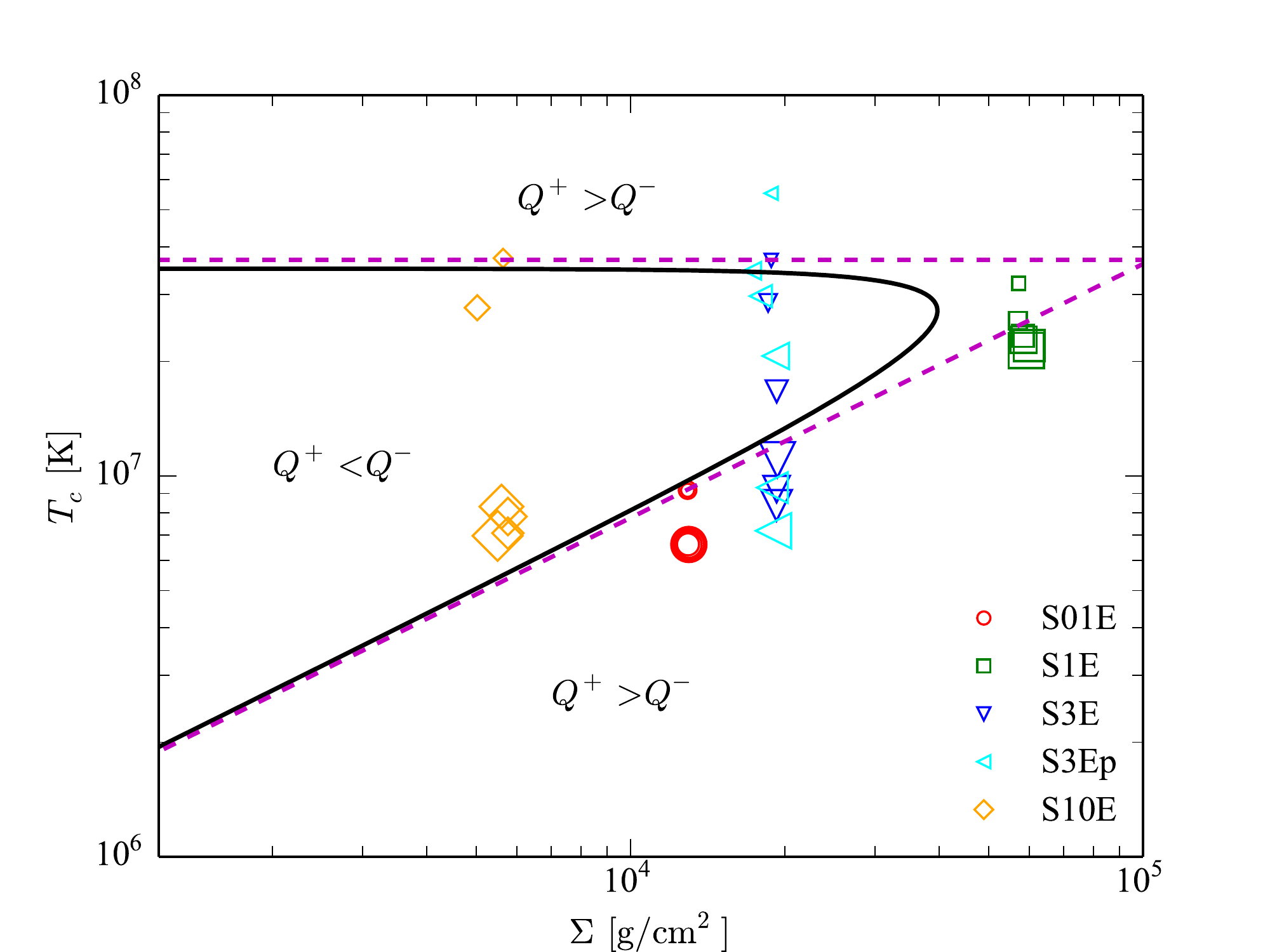} 
\includegraphics[width=0.48\columnwidth]{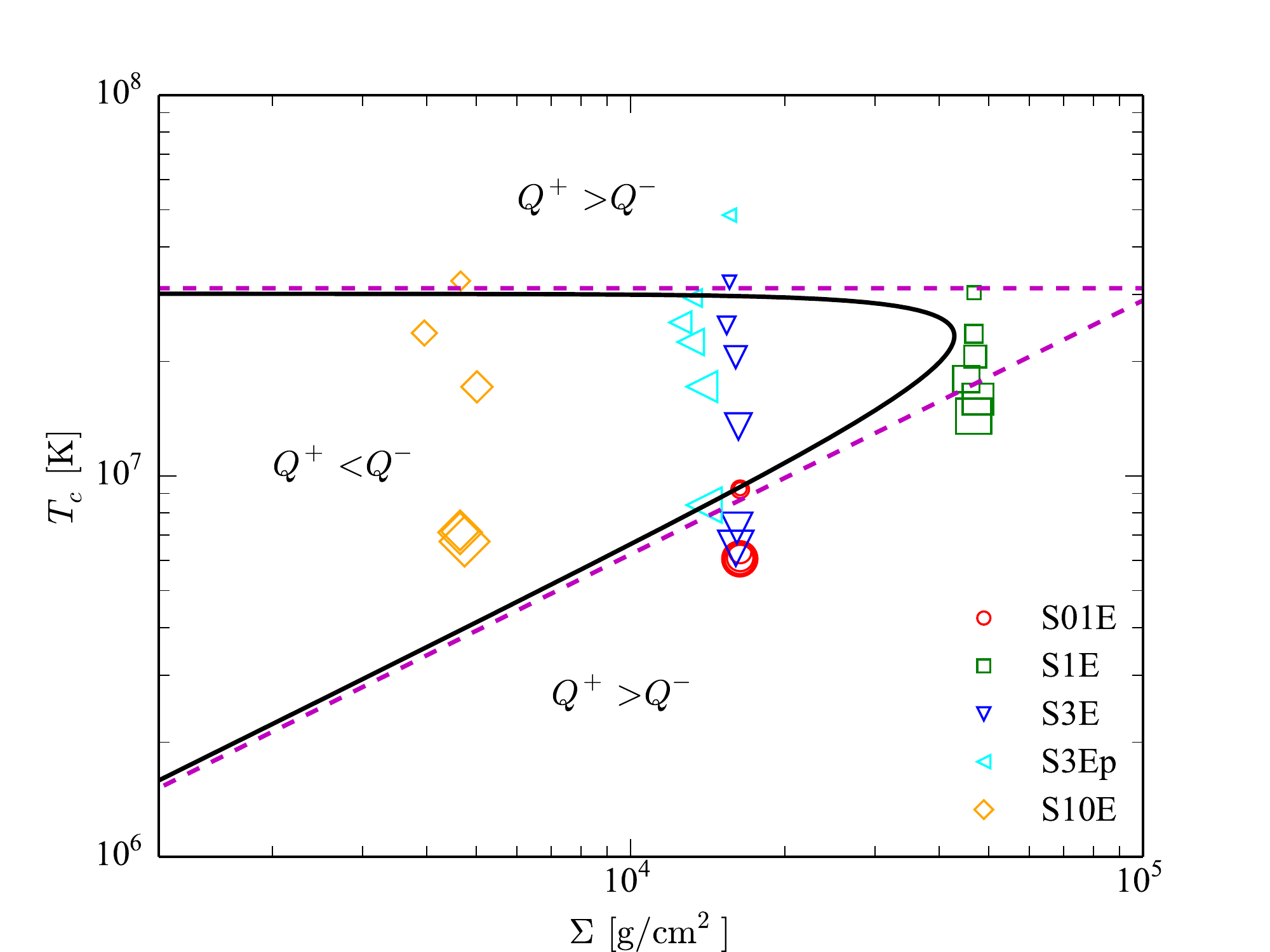} 
\caption{Thermal equilibrium ($T_c$-$\Sigma$) diagram for thin-disks at $R = 10 r_g$ (left panel) and $15 r_g$ (right panel). The solid line is for the standard Shakura-Sunyaev model with $\alpha = 0.02$, with the asymptotic gas- (bottom) and radiation- (top) pressure-dominated branches shown by dashed lines. The red circles show the evolution of simulation S01E, green squares the evolution of S1E, blue downward triangles the evolution of S3E, cyan leftward triangles the evolution of S3Ep, and yellow diamonds the evolution of S10E.  Data from the simulations have been time averaged over 3 ISCO orbital periods and radially averaged over intervals of 10 zones.  Increasing point sizes correspond to time intervals centered at $t = 0$, 5000, 10000, 15000, 20000, and $25000\,GM/c^3$, respectively. 
\label{fig:S-curve}}
\end{figure}

In this work, we consider five cases, which can be illustrated nicely by considering their positions in the $T_c$-$\Sigma$ plane, compared to an appropriate thermal equilibrium ($Q^+ = Q^-$) curve (Figure \ref{fig:S-curve}).  All five simulations assume a Schwarzschild black hole with $m = M_\mathrm{BH}/M_\odot = 6.62$. The viscosity parameter is $\alpha = 0.02$. The value of $M_\mathrm{BH}$ is chosen to make connection with earlier numerical work \citep[e.g.][]{Hirose09, Jiang13, Mishra16}, while the value of $\alpha$ is motivated by values typically seen in global MHD simulations of MRI turbulence with weak, local fields \citep[][and references therein]{Hawley11,Hawley13}. As can be seen in Figure \ref{fig:S-curve}, simulation S01E, with $\dot{m} = 0.01$, lies on the putative stable, gas-pressure-dominated branch, while simulation S1E, with $\dot{m} = 1$, lies near the transition between the stable and unstable branches, and S3E, S3Ep, and S10E, with $\dot{m}=3$, 3, and 10, respectively, lie on or near the unstable, radiation-pressure-dominated branch. Simulations S3E and S3Ep differ only in that the temperature is perturbed by a factor of 1.5 above its initial, equilibrium value in simulation S3Ep. Table \ref{tab:params} gives the full parameters for each simulation.

\section{Results}

\begin{figure}
\includegraphics[width=0.9\columnwidth]{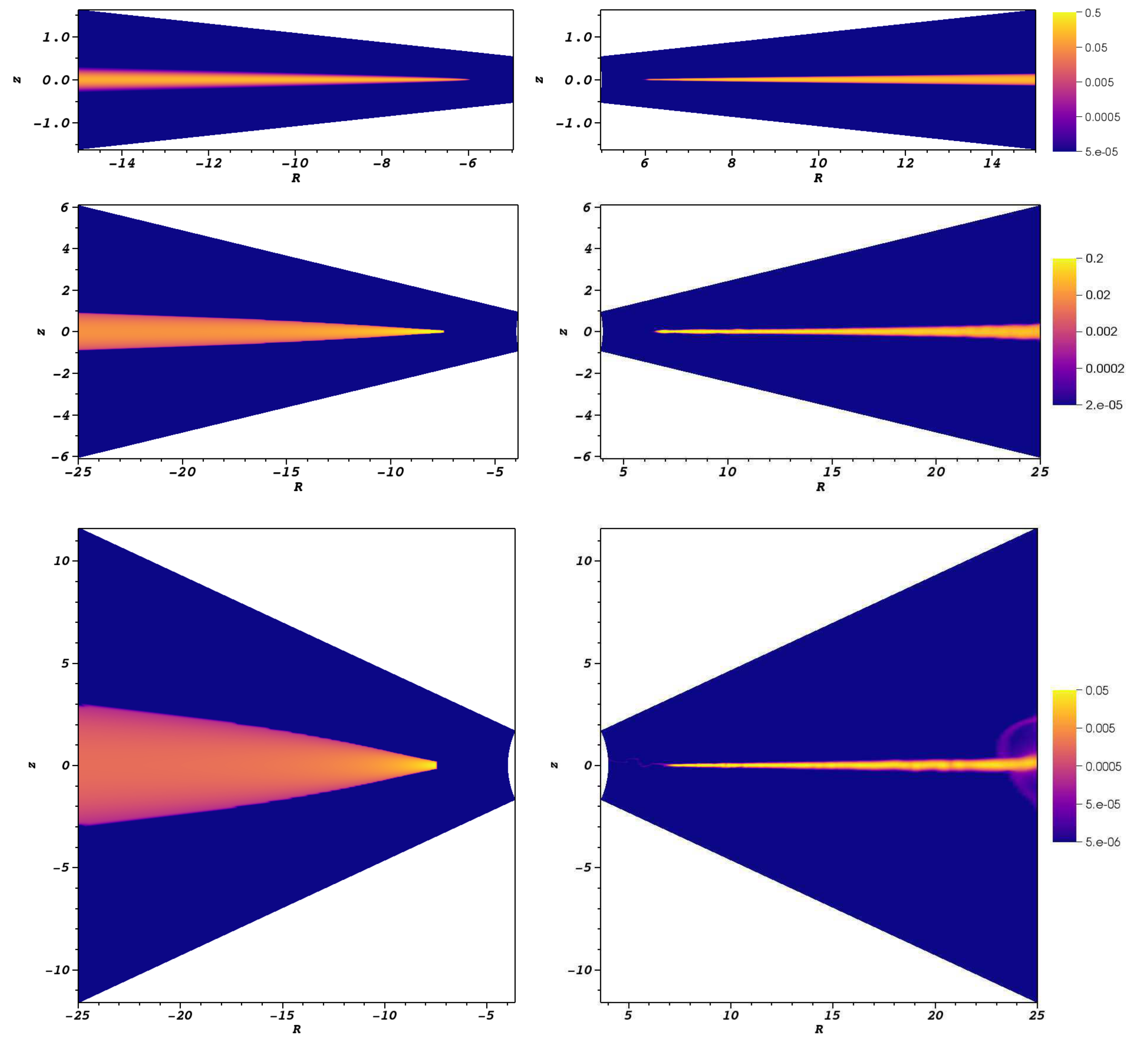} 
\caption{Pseudo-color plots of mass density (cgs units) for simulations S01E (top), S3E (middle), and S10E (bottom). The left- and right-hand panels correspond, respectively, to $t = 0$ and $t = 27475\,GM/c^3$. Note the vertical collapse of simulations S3E and S10E.
\label{fig:slices}}
\end{figure}

The Shakura-Sunyaev $\alpha$-viscosity model should, in general, only be studied on timescales longer than the viscous time, $t > t_\mathrm{vis} = r^2/\nu = r^2/(\alpha c_s H$), where the effects of turbulence can, in principle, be averaged over.  For thin disks with $\alpha = 0.02$, $t_\mathrm{vis} \sim 10^5\,GM/c^3 = 3\,$s at $10 r_g$, i.e., roughly of the same order as our simulations. For reference, the orbital period at the ISCO in these simulations is $92.3\,GM/c^3 = 0.003$ s, meaning all these simulations ran for hundreds of ISCO orbits. This is especially important when considering issues of stability, as simulations would need to run for at least this long to be deemed stable. In our case, many of our simulations showed clear signs of instability on considerably shorter timescales. For instance, of the simulations shown in Figure \ref{fig:slices}, only simulation S01E appears to have remained stable (i.e., close to its initial configuration) for its duration.  The other simulations, S3E and S10E in this case, by contrast, show clear evidence of vertical collapse, indicative of the thermal instability. This same conclusion is even more apparent in Figure \ref{fig:S-curve}, where we follow the evolutionary tracks of each simulation in the $T_c$-$\Sigma$ plane. While simulation S01E remains close to its initial location, most of the others drift a considerable distance in $T$, again as expected for a thermal instability. In the next few sections, we will explore each of these cases more thoroughly.

\subsection{Stable, Gas-Pressure-Dominated Disk}

\begin{figure}
\includegraphics[width=0.48\columnwidth]{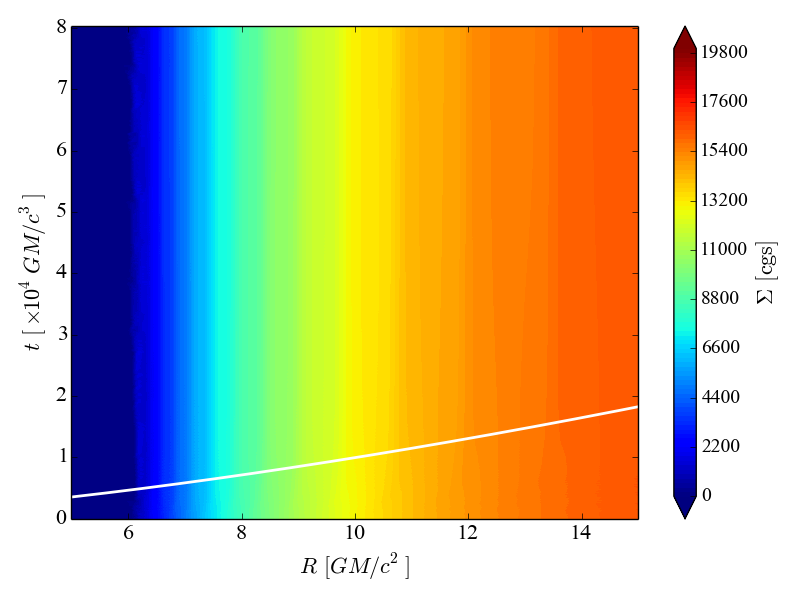}
\includegraphics[width=0.48\columnwidth]{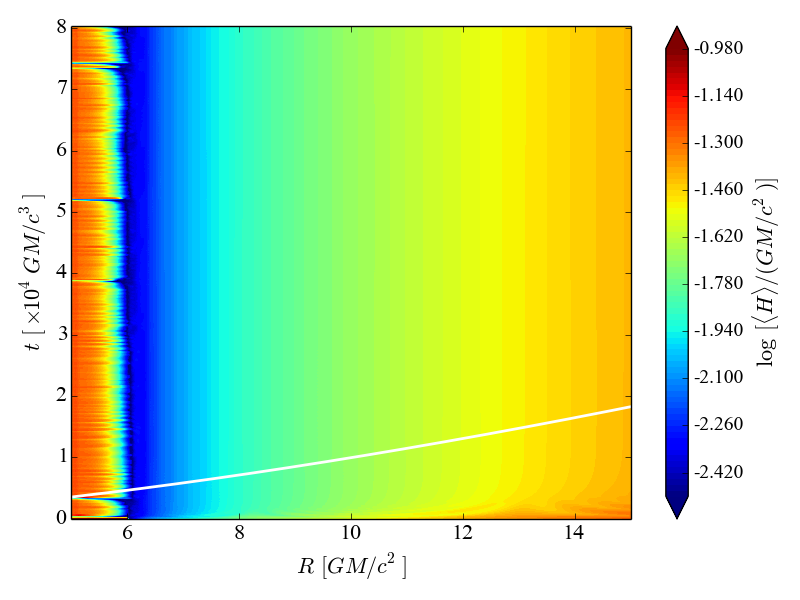} 
\caption{Space-time plots for the surface density, $\Sigma$, (left) and density-squared-weighted height, $\langle H \rangle$, (right) for simulation S01E. Solid curves show the estimated thermal time of an equilibrium disk, $2\pi/(\alpha \Omega)$. Notice there is very little evolution with time in this simulation.
\label{fig:S01E}}
\end{figure}

Before presenting further evidence that some of our simulations exhibit true instabilities, it may be helpful for us to demonstrate that our code can successfully evolve a stable, thin accretion disk, which was the main purpose of simulation S01E. This was the only gas-pressure-dominated (i.e., middle region) case that we considered. Figure \ref{fig:S01E} shows time histories of two of the key disk variables -- $\Sigma$ and $H$ -- for this simulation. The takeaway point is that, other than a brief initial adjustment, there is very little evolution in these variables, indicating stable evolution at close to the prescribed configuration. 

Although such thermal stability is the expected behavior for gas-pressure-dominated disks based upon the Shakura-Sunyaev model, there is observational evidence that this particular disk configuration may not manifest itself in nature. At the low accretion rate associated with this model ($\dot{m} \lesssim 0.02$), BHXRBs uniformly appear in the so-called ``Hard'' state \citep{Maccarone03,Remillard06}, which is usually interpreted as the inner part of the accretion flow being hot, thick, and advection-dominated \citep{Esin97,Done07}, very unlike the geometrically thin, Shakura-Sunyaev solution studied here. We can only speculate that additional physics that is not represented in our simulation (e.g. thermal conduction) may prevent this particular solution from being realized.

\subsection{Thermal Instability}
\label{sec:thermal}

\begin{figure}
\includegraphics[width=0.48\columnwidth]{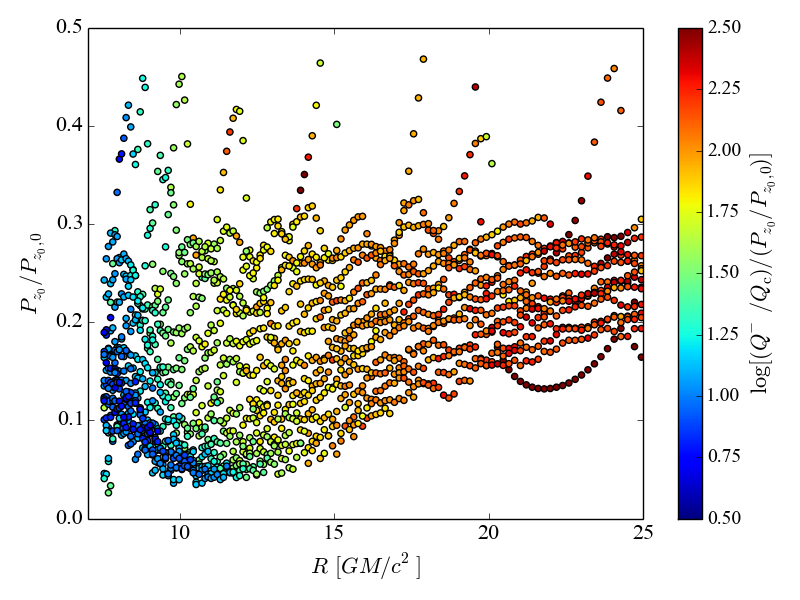} 
\includegraphics[width=0.48\columnwidth]{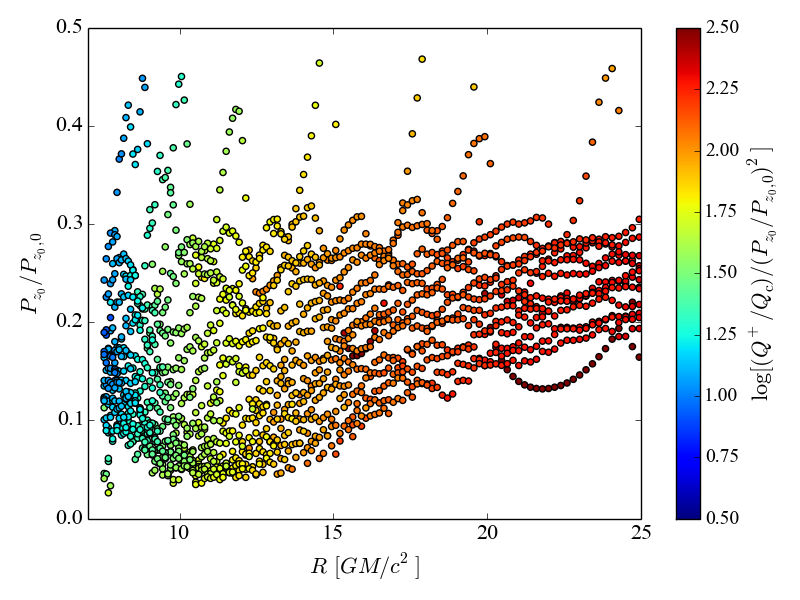} 
\caption{Scatter plots of the cooling rate, $Q^-$, normalized by $Q_c (P_{z_0}/P_{z_0,0})$ (left panel) and heating rate, $Q^+$, normalized by $Q_c (P_{z_0}/P_{z_0,0})^2$ (right panel), for simulation S3E, where $Q_c = c\Omega^2 H_0/\kappa^\mathrm{s}$, with $H_0 = 0.1\,GM/c^2$, $P_{z_0}$ is the total mid-plane pressure, and $P_{z_0,0}$ is the initial total mid-plane pressure. The near constant color of the scatter points at a given radius confirms that $Q^-$ and $Q^+$ follow the expected scalings with $P_{z_0}$. Data are sampled over the first $10^4\,GM/c^3$.
\label{fig:scalings}}
\end{figure}

The rest of the simulations we consider all lie on the radiation-pressure-dominated branch and are thus subject to various instabilities. We start by considering the thermal instability. To understand this instability, we note that the local cooling rate per unit area of the disk surface is proportional to the radiation energy density at the mid-plane, $a_\mathrm{R}T_\mathrm{c}^4$.  Since the opacity in these disks is dominated by electron scattering, the cooling should scale as $Q^- \propto T^4/\Sigma \propto P_{z_0}/\Sigma$.  The heating rate per unit area, on the other hand, is given by the vertically integrated stress times the rate of strain, or $Q^+ \sim H \tau_{R\phi} R \vert d\Omega/dR \vert$.  In the case where the disk is supported vertically predominantly by radiation pressure, the disk half thickness is proportional to the surface radiation flux, $H \propto Q^- \propto T^4/\Sigma$.  For a standard alpha disk, where $\tau_{R\phi} = \alpha P = \alpha a_\mathrm{R} T^4/3$, this means that the heating rate scales as $Q^+ \propto T^8/\Sigma \propto P_{z_0}^2/\Sigma$. Figure \ref{fig:scalings} demonstrates that our simulated radiation-pressure-dominated disks indeed follow the $Q^- \propto P_{z_0}$ and $Q^+ \propto P_{z_0}^2$ scalings. The result is that perturbative increases (decreases) in the midplane temperature should result in runaway heating (cooling).  Furthermore, the resulting instability should grow on the thermal timescale, $t_\mathrm{th} \sim (\alpha \Omega)^{-1}$.  

\begin{figure}
\includegraphics[width=0.48\columnwidth]{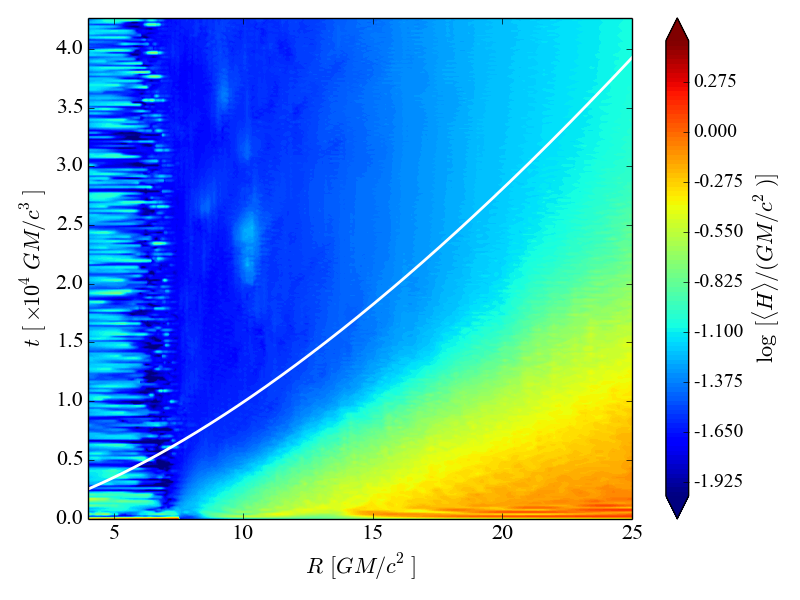} 
\includegraphics[width=0.48\columnwidth]{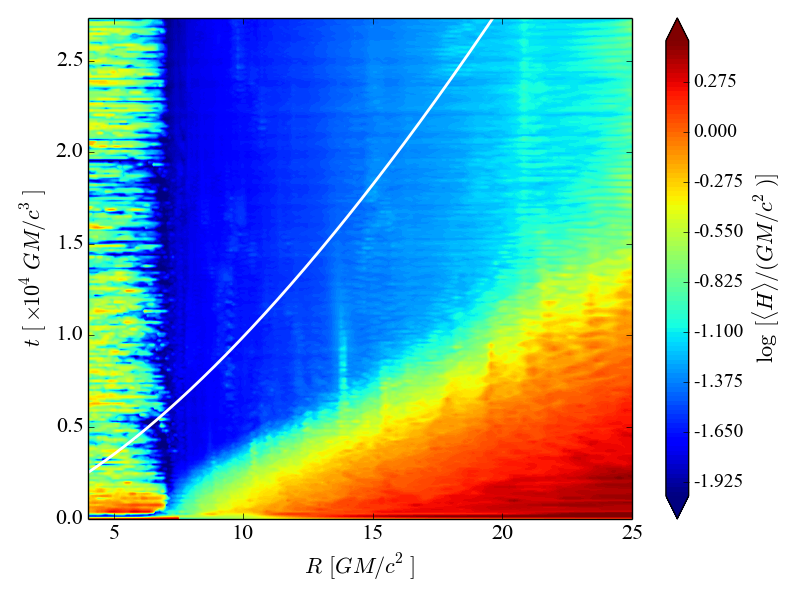} 
\caption{Space-time plot for the density-squared-weighted height of the disk in the S3E (left) and S10E (right) simulations. Solid curves show the estimated thermal time of an equilibrium disk, $2\pi/(\alpha \Omega)$.
\label{fig:height}}
\end{figure}

\begin{figure}
\includegraphics[width=0.48\columnwidth]{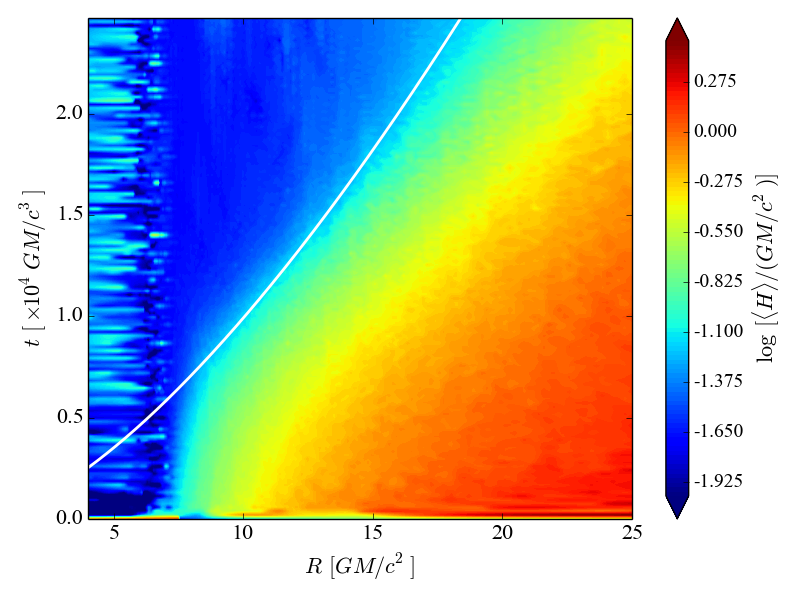} 
\includegraphics[width=0.48\columnwidth]{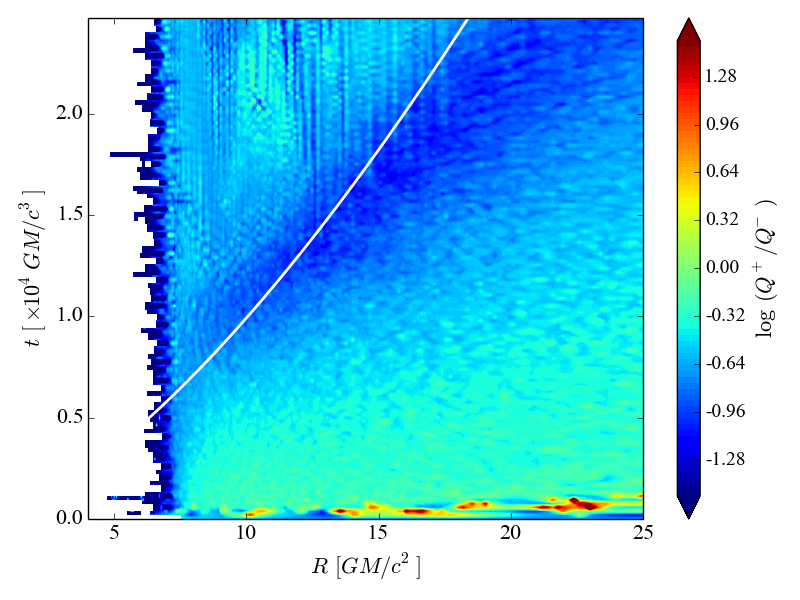} 
\caption{Space-time plot for the density-squared-weighted height of the disk (left) and the ratio of the heating rate, $Q^+$, to the cooling rate, $Q^-$, (right) for the S3Ep simulation. Solid curves show the estimated thermal time of an equilibrium disk, $2\pi/(\alpha \Omega)$.
\label{fig:S3Ep}}
\end{figure}

Figure \ref{fig:height}, which shows the density-squared-weighted height, 
\begin{equation}
<H> = \sqrt{\frac{\int^{\theta_\mathrm{max}}_{\theta_\mathrm{min}} \sqrt{-g}\rho^2 (\theta - \pi/2)^2 d\theta}{\int^{\theta_\mathrm{max}}_{\theta_\mathrm{min}} \sqrt{-g}\rho^2 d\theta}}~,
\end{equation}
for simulations S3E and S10E, clearly points to a thermal collapse (follow a vertical line at any radius and notice the height decreasing with time until it stabilizes at a new value). From the figure, it appears that the collapse happens slightly faster than the local thermal time (indicated by the solid, white line), which was also seen in \citet{Mishra16}. Simulation S3Ep, the perturbed case, initially shows a modest growth in height, but still ultimately collapses on roughly the thermal timescale (left panel of Figure \ref{fig:S3Ep}). The collapse of simulation S1E is not as dramatic as these others, as it represents an intermediate case. 

\begin{figure}
\includegraphics[width=0.48\columnwidth]{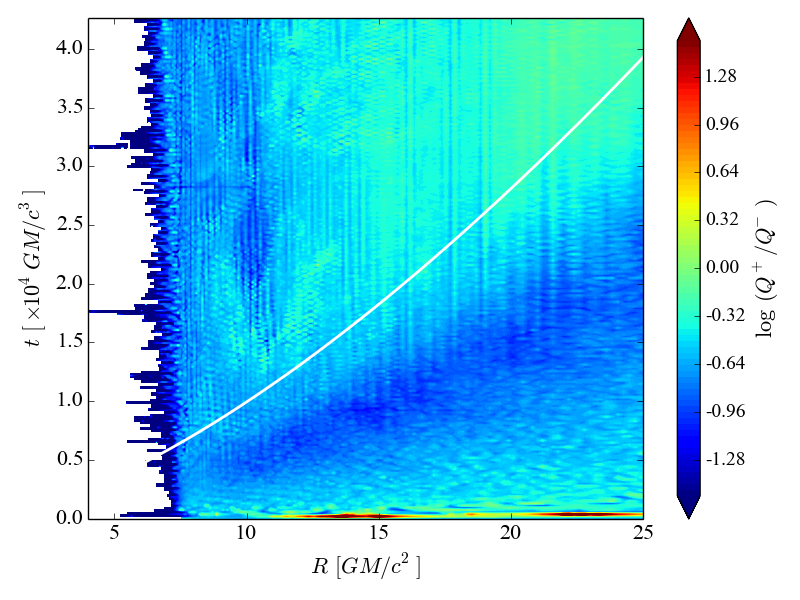} 
\includegraphics[width=0.48\columnwidth]{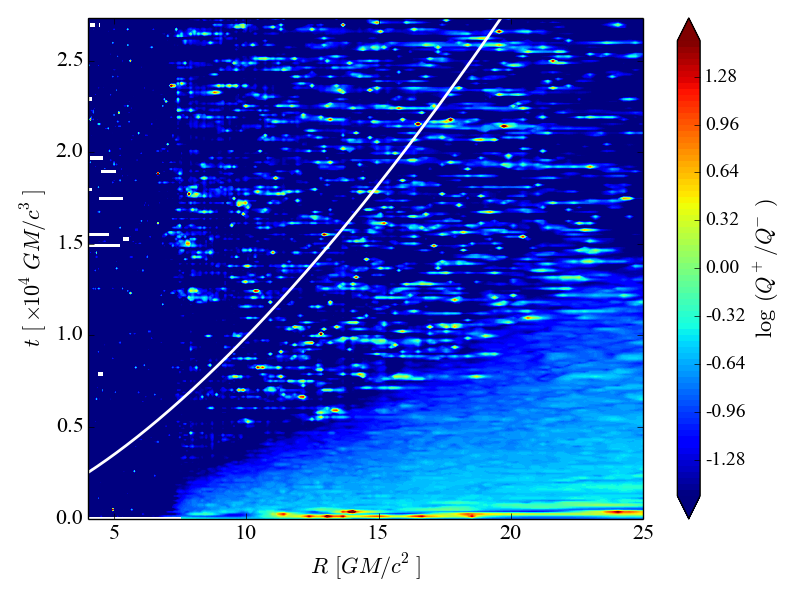} 
\caption{Space-time plot for the ratio of the heating rate, $Q^+$, to the cooling rate, $Q^-$, for the S3E (left) and S10E (right) simulations. Solid curves show the estimated thermal time of an equilibrium disk, $2\pi/(\alpha \Omega)$. Green shades indicate thermal equilibrium ($Q^+ = Q^-$).
\label{fig:heating_cooling}}
\end{figure}

Figure \ref{fig:heating_cooling} confirms that the collapse seen in Figure \ref{fig:height} happens because cooling (blue shades) rather quickly exceeds heating (red shades) in the disk. The process seems to accelerate as the disks collapse. At later times, simulation S3E appears to reach a new thermal equilibrium (green shades), at least in part of the disk ($20 \le r/r_g \le 25$). This is consistent with Figure \ref{fig:S-curve}, where it appears that this simulation collapses down to the lower stable (gas-pressure-dominated) branch and settles into a new solution there. Similar transitions between the unstable to stable branches were suggested in some earlier shearing box studies \citep{Agol01,Turner02}. Simulation S10E, on the other hand, never regains its thermal equilibrium, not quite reaching the lower branch in Figure \ref{fig:S-curve}. We suspect that this is simply the result of the simulation not having sufficient resolution to follow the collapse further. We expect that with additional resolution this simulation would continue to collapse until it reached the stable branch and established a new thermal equilibrium.

It is important to note the thermal {\em collapse} (i.e., runaway cooling) of {\it all three} simulations (S3E, S3Ep, and S10E), especially S3Ep. For that simulation, we purposefully perturbed the initial disk solution to a higher temperature to put it above the thermal equilibrium curve (see Figure \ref{fig:S-curve}). According to our scaling arguments above, this {\em should} have resulted in runaway heating and thermal {\em expansion} of the disk. Figure \ref{fig:S3Ep} shows that this was the case for a very brief period of time. Ultimately, however, this simulation followed an evolutionary track very similar to the unperturbed simulation S3E, although on a somewhat delayed timescale. This has been a common occurrence in numerical simulations of thermally-unstable, radiation-pressure-dominated disks \citep{Jiang13,Mishra16}, i.e., they tend to exhibit thermal collapse preferentially over thermal expansion with only a few exceptions \citep[e.g. Fig. 1 of][]{Jiang13}. In our case, one thing to consider is that our disks do not actually start in hydrostatic equilibrium, even though we initiate them as close as possible to the Shakura-Sunyaev solution. The initial setup may, in some way, favor cooling over heating to such a degree that all of the simulations end in collapse. However, the reason for this apparent preference, and even whether it is physical or numerical, remains unclear at this time.

The longer term evolution of these disks would of course be of interest. As the new solutions have lower $\dot{m}$ than what is being fed in from larger radii, mass must be accumulating somewhere in the disk, as is indeed happening near the outer boundary of each simulation. This could potentially lead to some sort of limit-cycle behavior, where the disk switches between low- and high-$\dot{m}$ solutions \citep{Janiuk02,Ohsuga06}. This possibility will be explored in future work.

In this paper we are only considering radiation hydrodynamic simulations, i.e., the role of magnetic fields is neglected, other than that they are presumed to be the underlying source of the stresses that we model as a viscosity. However, if strong magnetic fields are present, they may, in principle, stabilize radiation-pressure-dominated disks against the thermal instability seen in our work \citep{Begelman07,Oda09,Sadowski16}. Even with weaker magnetic fields, the cooling of the disk may be altered by magnetic buoyancy.

\subsection{Viscous Instability}
\label{sec:viscous}

\begin{figure}
\includegraphics[width=0.48\columnwidth]{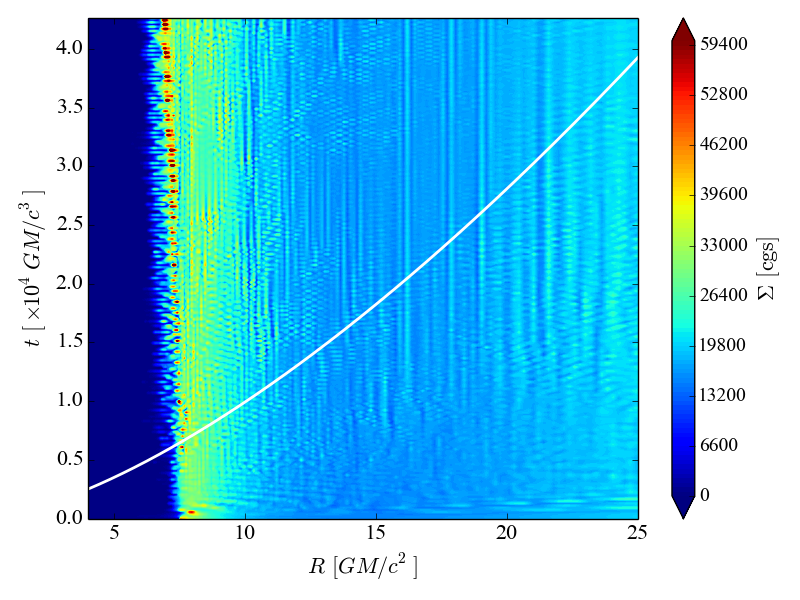}
\includegraphics[width=0.48\columnwidth]{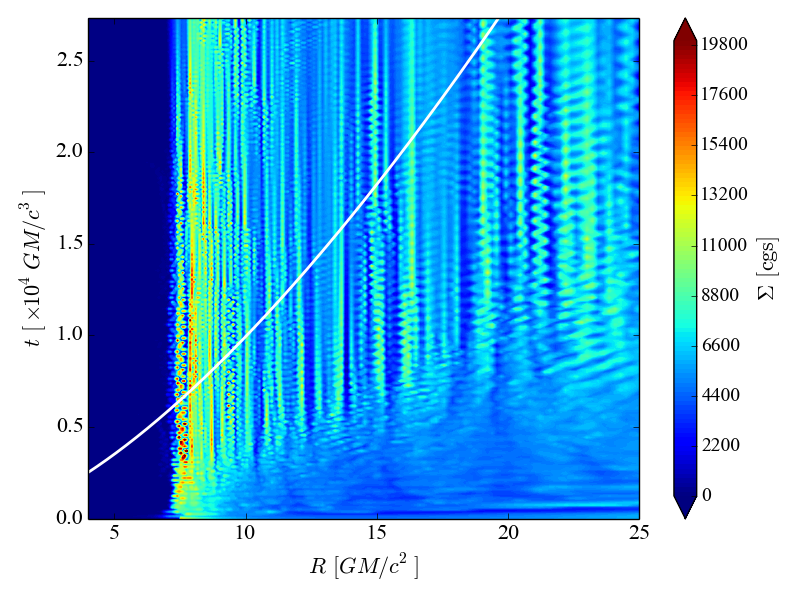} 
\caption{Spacetime plot for the surface density, $\Sigma$, of the disks in the S3E (left) and S10E (right) simulations. Solid curves show the estimated thermal time of an equilibrium disk, $2\pi/(\alpha \Omega)$, which is approximately the growth time of the viscous instability. The extended vertical tracks indicate that these disks have broken up into rings.
\label{fig:sigma}}
\end{figure}

In the radiation-pressure-dominated regime, the vertically integrated stress, $W_{R\phi}$, in a standard Shakura-Sunyaev disk can be shown to be inversely proportional to the surface density, $\Sigma$.  This means that the evolution equation for $\Sigma$, which amounts to a nonlinear diffusion equation,
\begin{equation}
\frac{\partial\Sigma}{\partial t} \propto \frac{\partial}{\partial r} \left[ r^2 W_{R\phi} \right] ~,
\end{equation}
has a negative effective diffusion coefficient \citep{Lightman74}.  This implies that the disk should break up into high and low surface density rings with $\Delta R \gtrsim H$ on a timescale $t_\mathrm{LE} \sim (\Delta R/R)^2 t_\mathrm{vis}$, which turns out to be roughly the same as the thermal timescale, $t_\mathrm{th}$. We indeed see a breaking of our disks into rings, as exhibited by the long, vertical streaks of surface density in the spacetime diagrams of Figure \ref{fig:sigma}. 

\begin{figure}
\includegraphics[width=0.48\columnwidth]{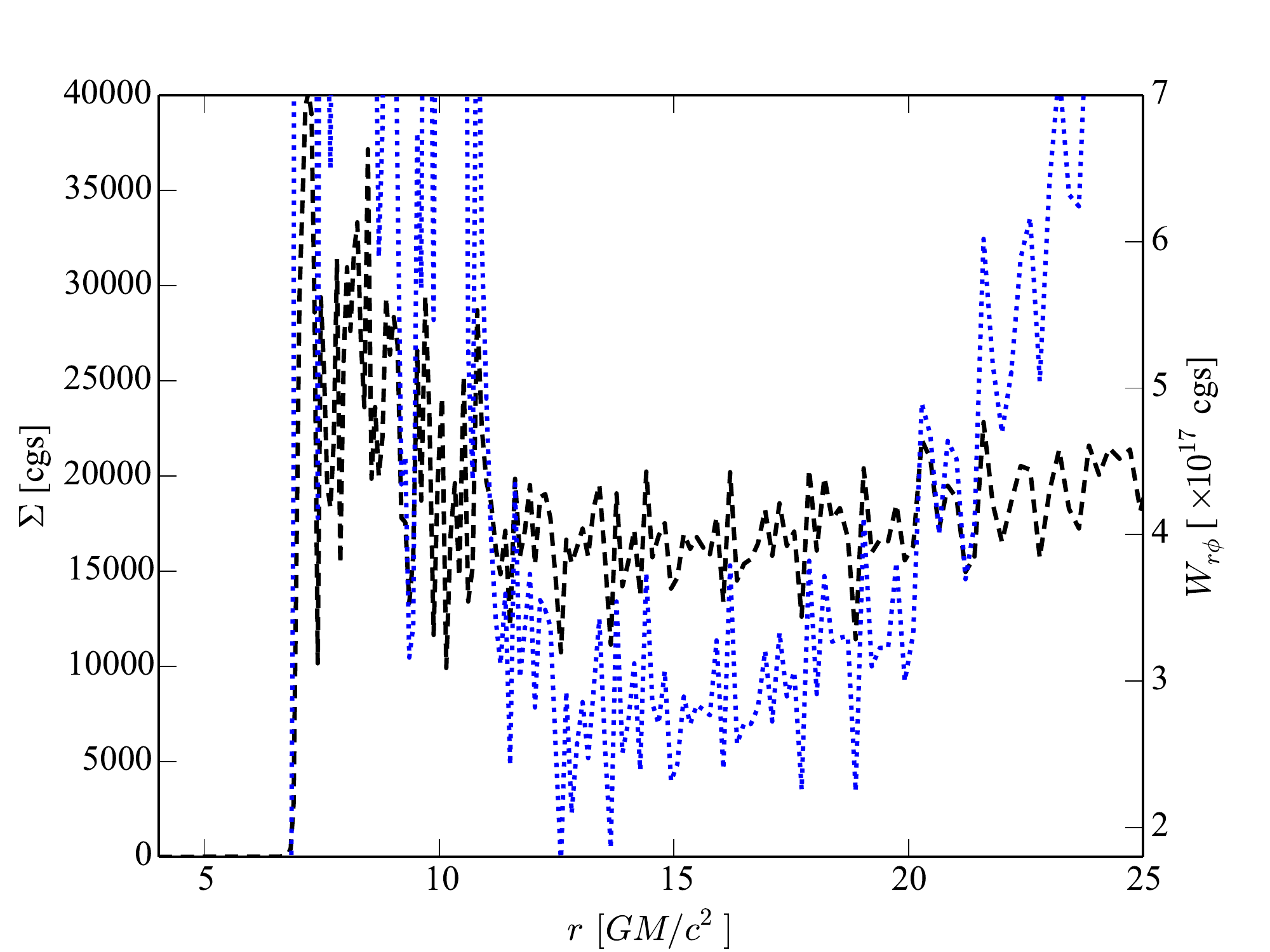}
\includegraphics[width=0.48\columnwidth]{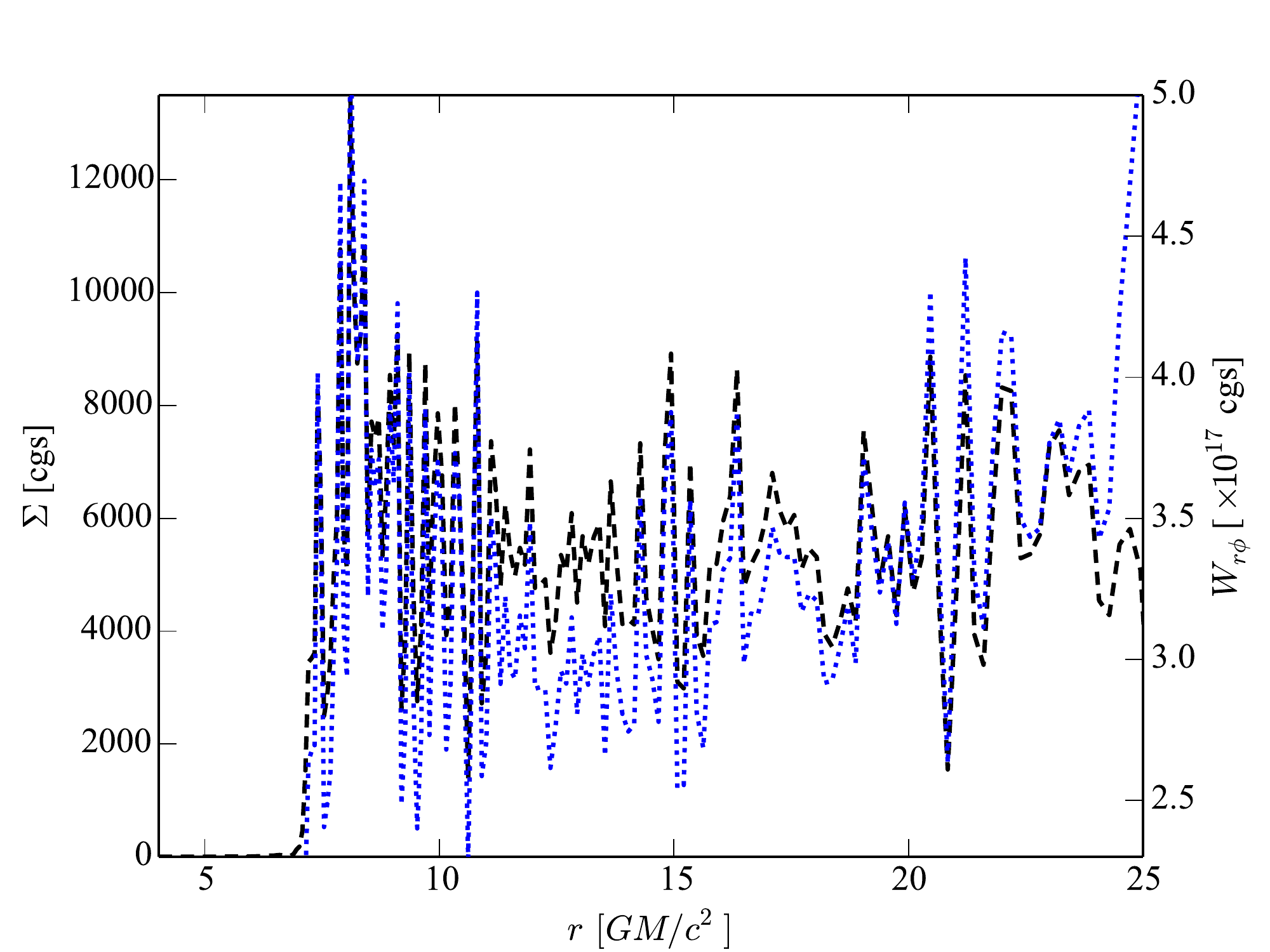} 
\caption{Lineout plots of the surface density, $\Sigma$, (black, dashed curve) and the vertically integrated stress, $W_{r\phi} = 2H\alpha P$, (blue, dotted curve) of the disk in the S3E (left) and S10E (right) simulations at $t = 22152\,GM/c^3$. Note the strong {\em positive} correlation between peaks and valleys in the two quantities, opposite of what would be expected for a viscous instability.
\label{fig:stress}}
\end{figure}

Although the disks in both simulations 3E and 10E break up into distinct radial rings with $\Delta R \sim H$ on roughly the local thermal timescale (solid line), as predicted by \citet{Lightman74}, we do not believe this is the same instability identified in that paper.  Our main reason for arguing this is that, as shown in Figure \ref{fig:stress}, the vertically integrated stress, $W_{R\phi}$, is actually high in regions where $\Sigma$ is high and low where $\Sigma$ is low, exactly opposite what is predicted by the viscous instability.

\begin{figure}
\includegraphics[width=0.48\columnwidth]{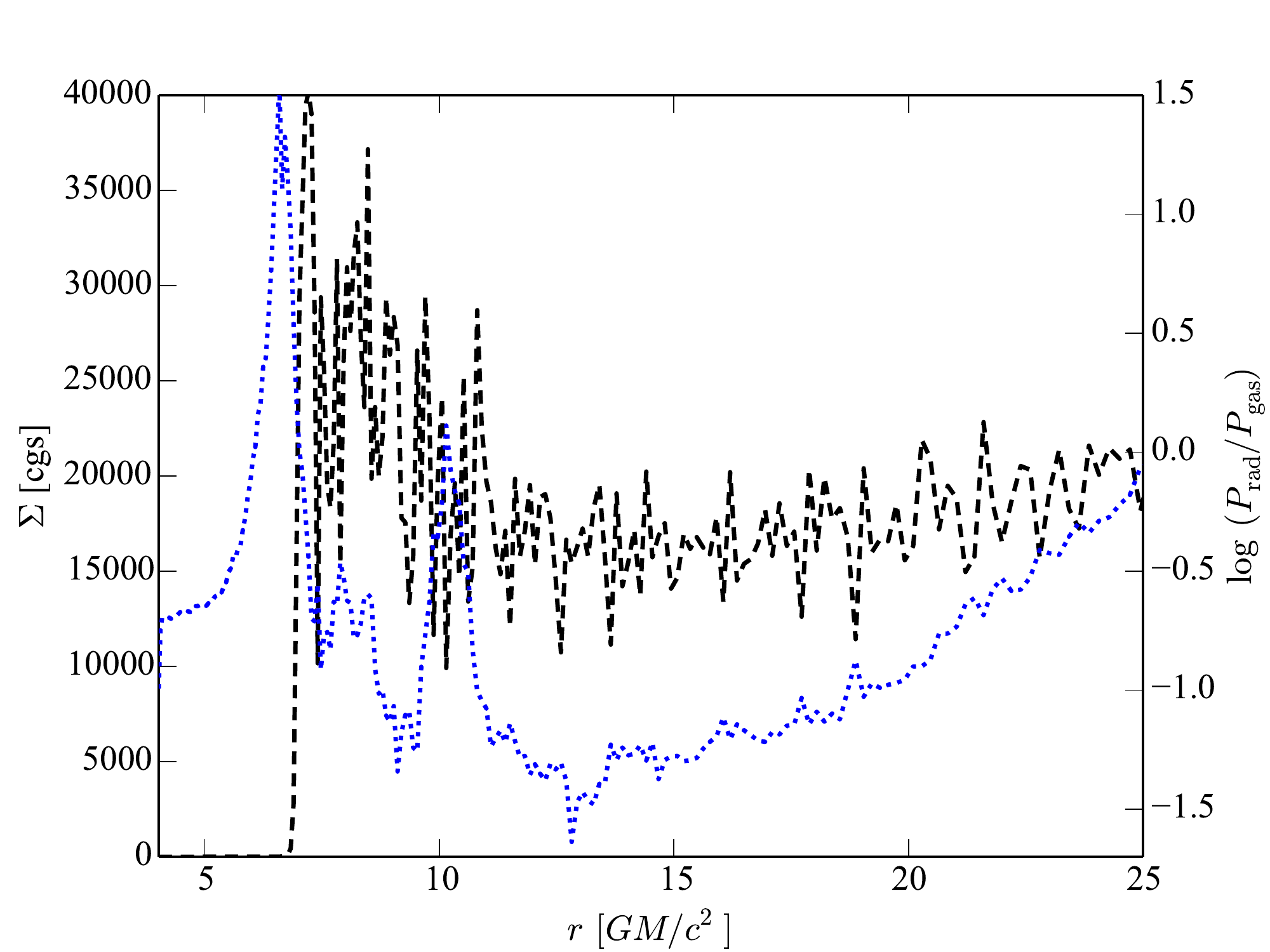}
\includegraphics[width=0.48\columnwidth]{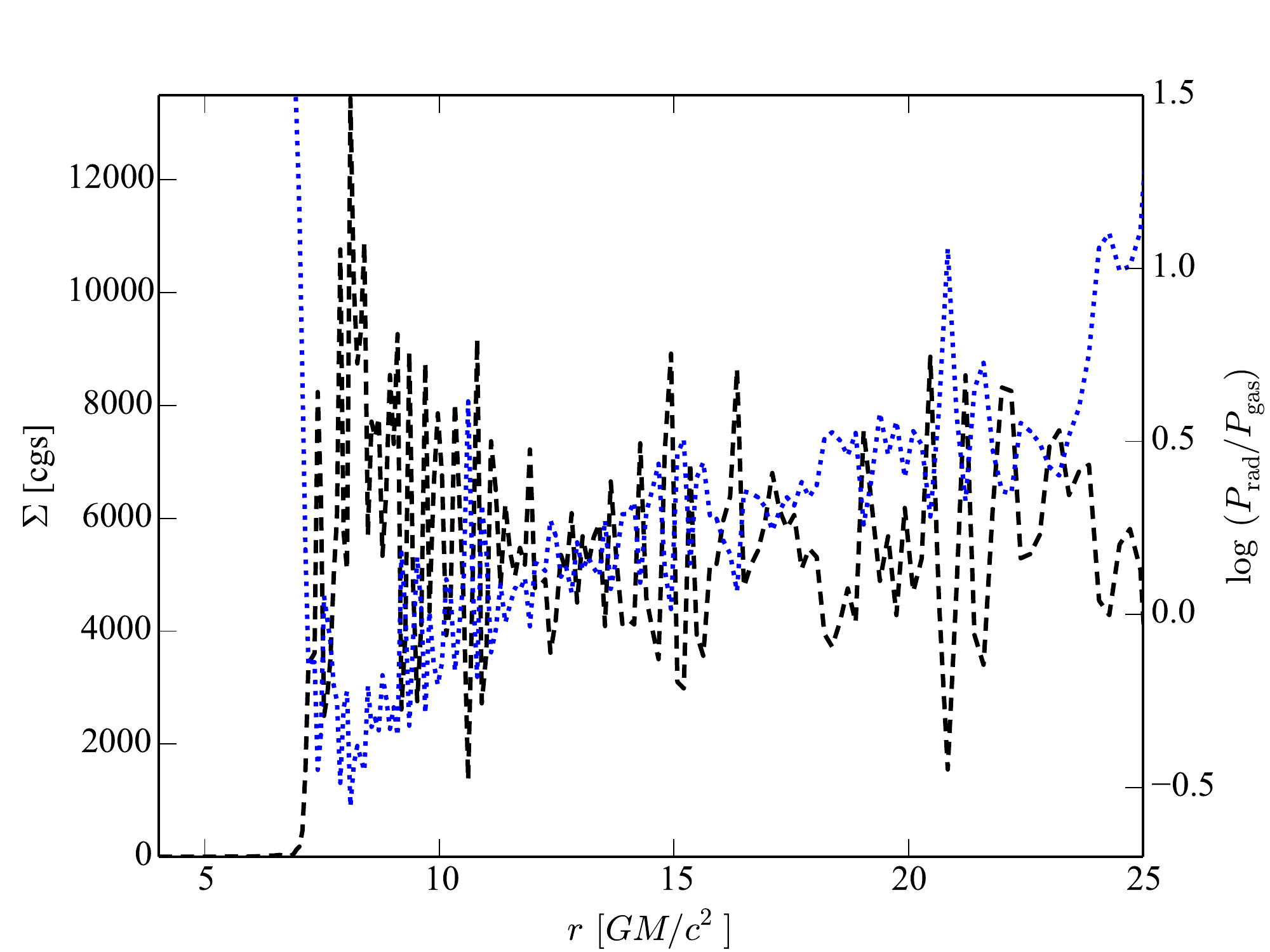} 
\caption{Lineout plots of the surface density, $\Sigma$, (black, dashed curve) and the density-weighted, shell-averaged ratio of radiation pressure to gas pressure, $P_\mathrm{rad}/P_\mathrm{gas}$, (blue, dotted curve) for simulations S3E (left) and S10E (right). Note the strong {\em negative} correlation between peaks and valleys in the two quantities.
\label{fig:prad_pgas}}
\end{figure}

\begin{figure}
\includegraphics[width=0.48\columnwidth]{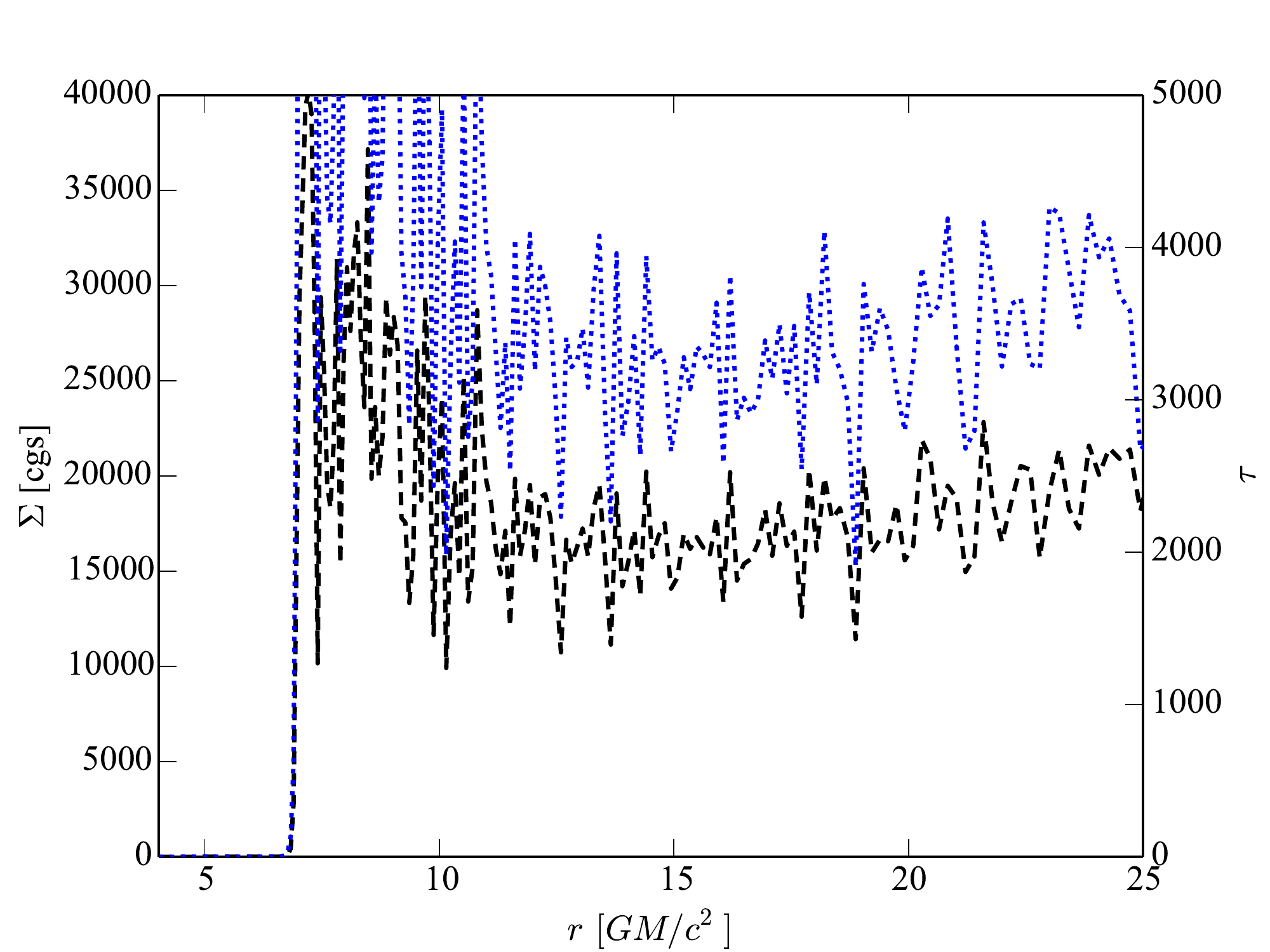}
\includegraphics[width=0.48\columnwidth]{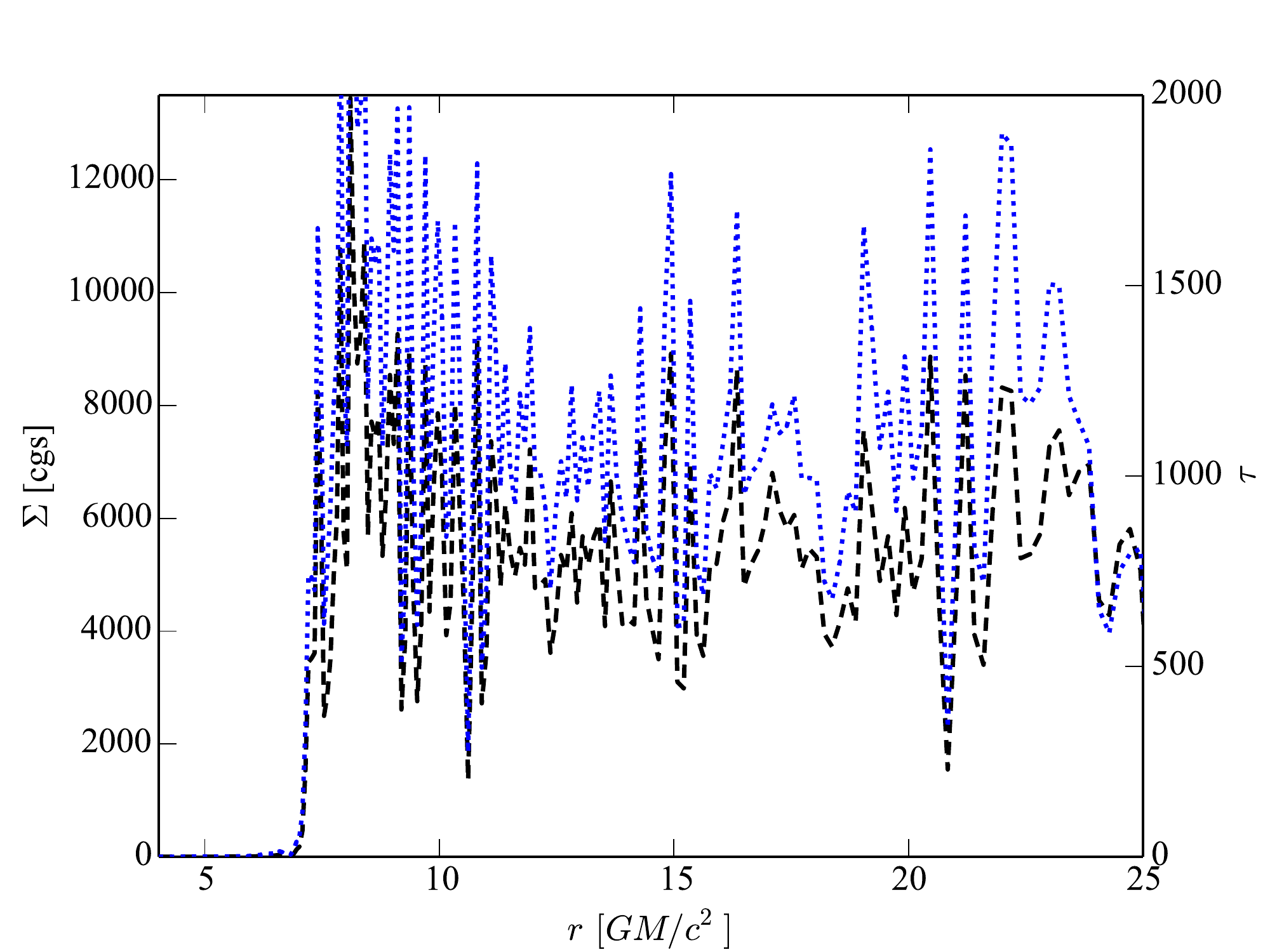} 
\caption{Lineout plots of the surface density, $\Sigma$, (black, dashed curve) and the density-weighted, shell-averaged ratio of radiation pressure to gas pressure, $P_\mathrm{rad}/P_\mathrm{gas}$, (blue, dotted curve) for simulations S3E (left) and S10E (right). Note the strong {\em positive} correlation between peaks and valleys in the two quantities, showing that low surface density gaps are also low optical depth gaps.
\label{fig:opacity}}
\end{figure}

The first clue as to what may, in fact, be driving our simulated disks to break into rings is to note that it only happens in the radiation-pressure-dominated cases (compare $\Sigma$ for simulation S01E in Figure \ref{fig:S01E} to $\Sigma$ for simulations S3E and S10E in Figure \ref{fig:sigma}). In fact, the low-density gaps in the disk appear to form where there are local peaks in the ratio of radiation-to-gas pressure, as shown in Figure \ref{fig:prad_pgas}. Also consistent with this idea is the fact that the opacity is notably lower in the low-density gaps, as shown in Figure \ref{fig:opacity}, providing the radiation an easier path to escape from the disk. This is expected behavior for a radiation-pressure-dominated medium; once low density channels form, the radiation will naturally follow these channels and thus reinforce the perturbations. The only question is, what allows the initial density inhomogeneities to form? In this case, it appears it may be convection. We find that our radiation-pressure-dominated disks are convectively unstable according to the Schwarzschild condition
\begin{equation}
\frac{dT}{dz} < \left(1-\frac{1}{\Gamma}\right) \frac{T}{P} \frac{dP}{dz} ~,
\end{equation}
whereas our gas-pressure-dominated case (S01E) is stable by the same criterion.

\subsection{Luminosity \& Variability}
\label{sec:luminosity}

\begin{figure}
\includegraphics[width=0.48\columnwidth]{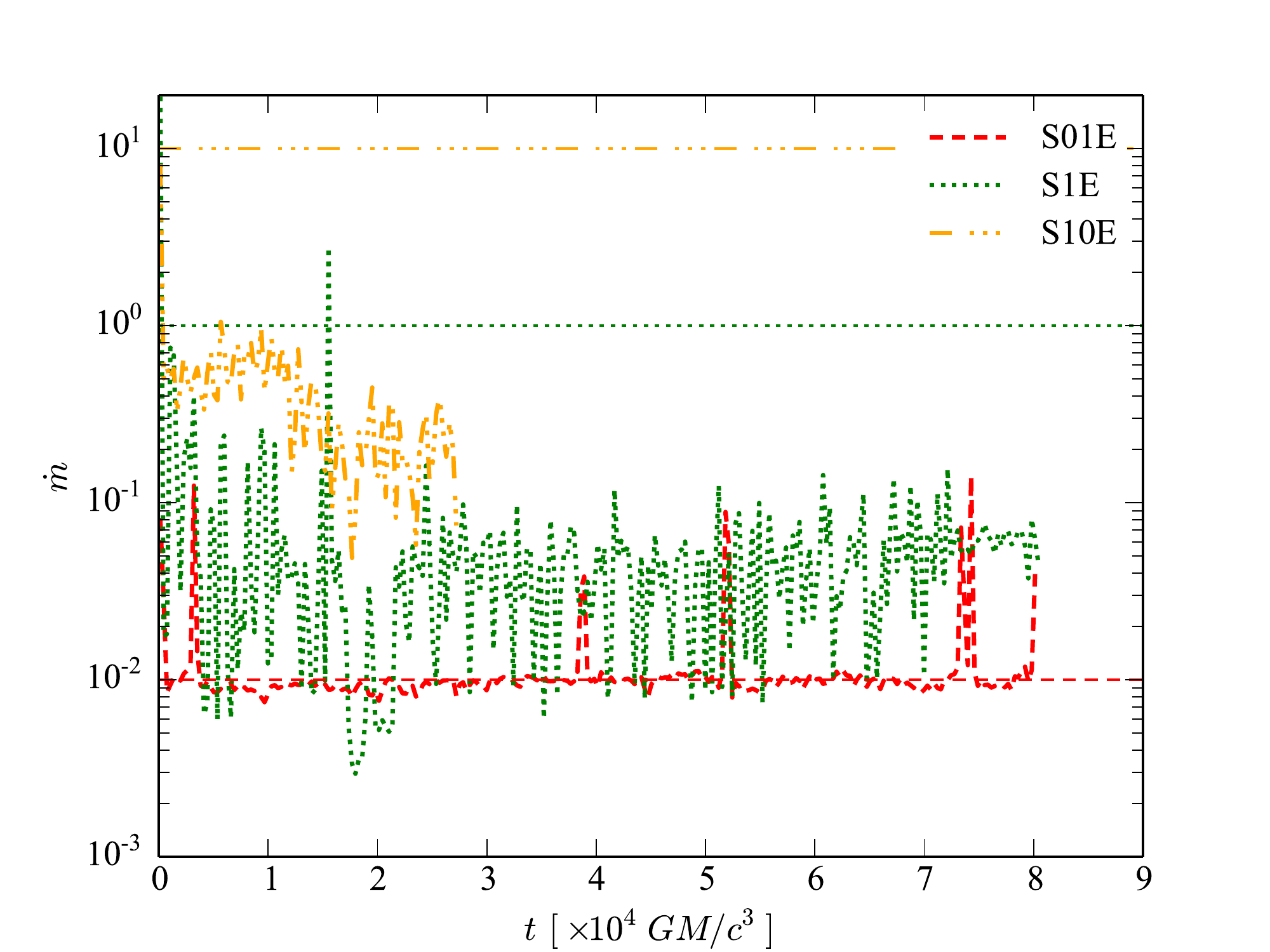} 
\caption{Mass accretion rate measured near the inner radial boundary of the simulation domain, scaled to Eddington, i.e., $\dot{m} = \dot{M}/\dot{M}_\mathrm{Edd}$. The thin lines show the target values of $\dot{m}$ for each simulation.
\label{fig:massFlux}}
\end{figure}

Because these are global, {\em radiation} hydrodynamic simulations, we can directly extract light curves and compare the measured luminosity with predictions for thin accretion disks. First, let us look at the measured accretion rates of the simulations, as it is mass accretion that ultimately powers the luminosity. Not surprisingly, given the instabilities we have already identified, the measured mass accretion rates often do not match the input value for that simulation. This is because, as the thermally unstable solutions collapse and seek new, stable equilibria, the mass accretion rate must necessarily drop. This follows from the fact that, for a given $R$, the stable, gas-pressure-dominated branch has a lower $\dot{m}$ than the unstable, radiation-pressure-dominated one. Figure \ref{fig:massFlux} shows the $\dot{m} = \dot{M}/\dot{M}_\mathrm{Edd}$ values measured near the inner boundaries of each simulation domain, where
\begin{equation}
\dot{M} =  -2\pi \int \sqrt{-g} \rho u^r \mathrm{d}\theta ~.
\end{equation}
For clarity, we only show the mass accretion histories of three of our simulations: S01E, S1E, and S10E. Note that simulation S01E accretes at exactly its targeted accretion rate ($\dot{m} = 0.01$), as expected for a stable, gas-pressure-dominated case, whereas simulations S1E and S10E accrete at significantly below their target rates, consistent with their thermal collapse. Simulations S3E and S3Ep (not shown) also accrete at significantly below their target levels. The thermally unstable simulations also exhibit far greater variability in their mass accretion rates than the stable S01E case.

\begin{figure}
\includegraphics[width=0.48\columnwidth]{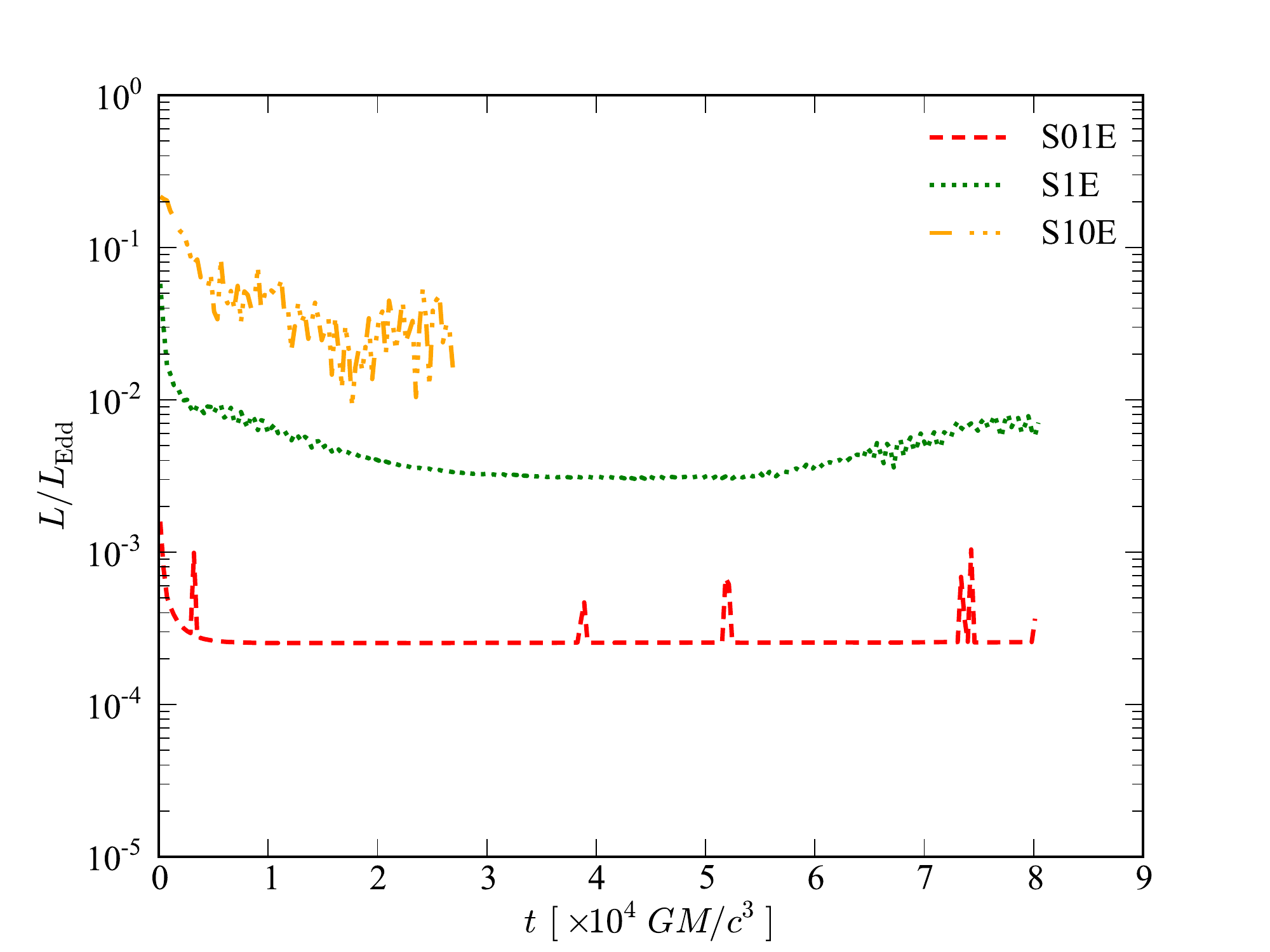} 
\includegraphics[width=0.48\columnwidth]{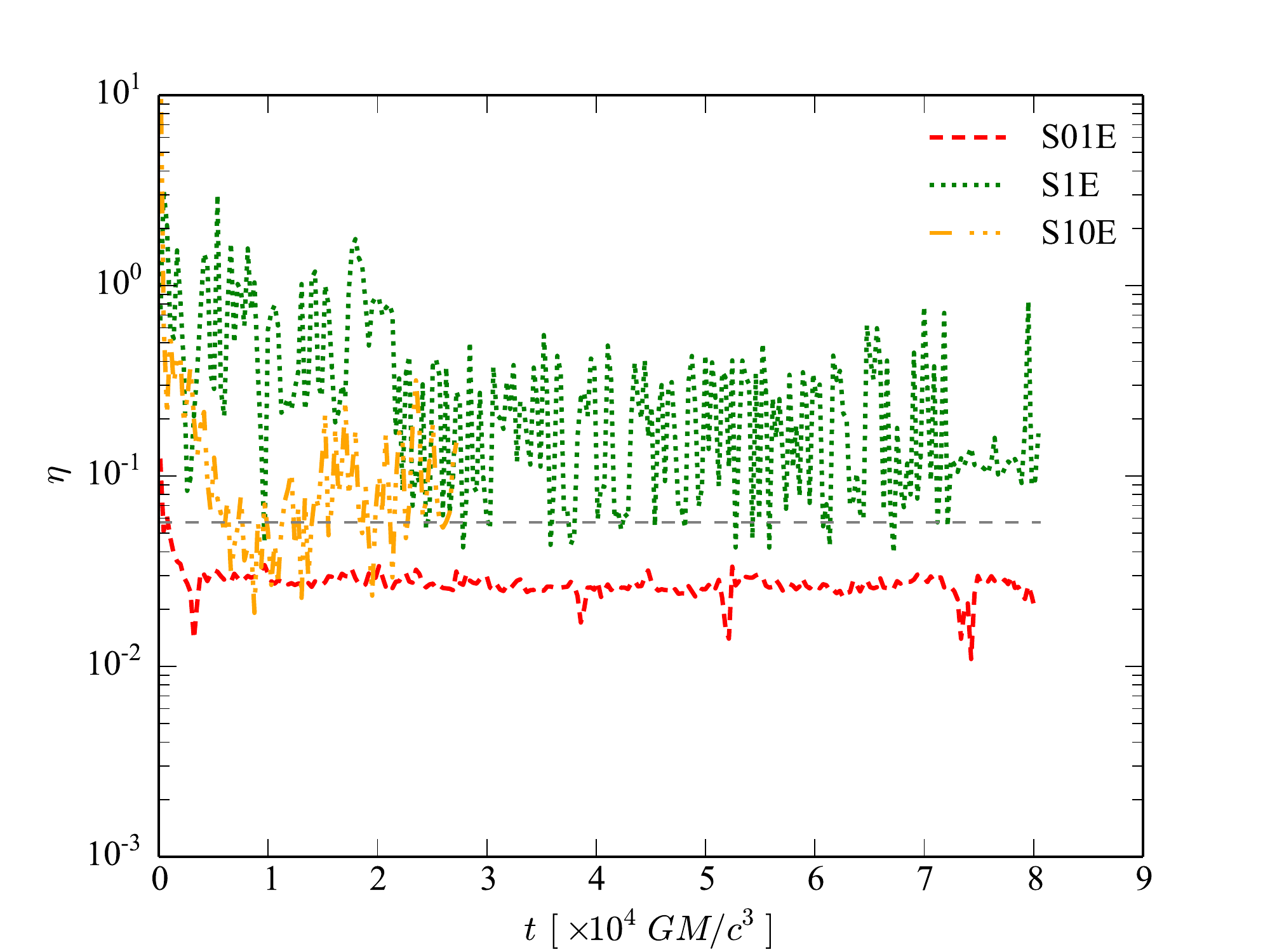} 
\caption{Left: Luminosity, summed over the top and bottom poloidal boundaries and an intermediate radial boundary (to exclude numerical artifacts at the outer radial boundary) of the simulation domain, scaled to Eddington. Right: Radiative efficiency, $\eta = L/\dot{M}c^2 = (L/L_\mathrm{Edd})/\dot{m}$, for each simulation, which can be compared to the expected value of $\eta = 0.057$ (gray, dashed line). Note that for these plots we employ moving averages equal to 3 ISCO orbital periods to smooth the data. 
\label{fig:luminosity}}
\end{figure}

Next we consider how efficiently this matter accretion is converted into radiation. To calculate the luminosity, we integrate 
\begin{equation}
L = -\int_S \sqrt{-g} R^\theta_t \mathrm{d}A_\theta ~,
\end{equation}
where $R^\theta_t$ is the radiative flux in the $\theta$-direction, normal to an area element $\mathrm{d}A_\theta$. For this work, we take the surface to be at constant $\theta$ near the top and bottom boundaries of each grid, out to a radius of $r = 15 r_g$. We are obviously underestimating the total luminosity as we are ignoring any radiation originating outside $15 r_g$, as well as any radiation passing through the inner and outer radial boundaries of our defined volume. However, both contributions should be relatively minor in terms of their contribution to the luminosity that would be measured at infinity. Indeed, we find that the radiated luminosity (Figure \ref{fig:luminosity}, left panel) closely matches the theoretical expectations corresponding to the observed mass accretion rates in Figure \ref{fig:massFlux}. For a Schwarzschild (non-rotating) black hole, the radiative efficiency, $\eta = L/\dot{M}c^2 = (L/L_\mathrm{Edd})/\dot{m}$, should be approximately 6\%. In the right panel of Figure \ref{fig:luminosity}, we plot the measured radiative efficiencies from some of our simulations as a function of time. Clearly, for simulations S1E and S10E, the efficiency varies considerably, but for the simulations that do find a stable solution, the value of $\eta$ is generally within a factor of a couple of its predicted value. Keep in mind, too, that because $\dot{m}$ is a measure of the instantaneous mass accretion rate at the inner boundary while the luminosity is measured over a much larger region, the two are not expected to vary on the same timescale, so $\eta$ should be expected to deviate significantly on short timescales.

\begin{figure}
\includegraphics[width=0.48\columnwidth]{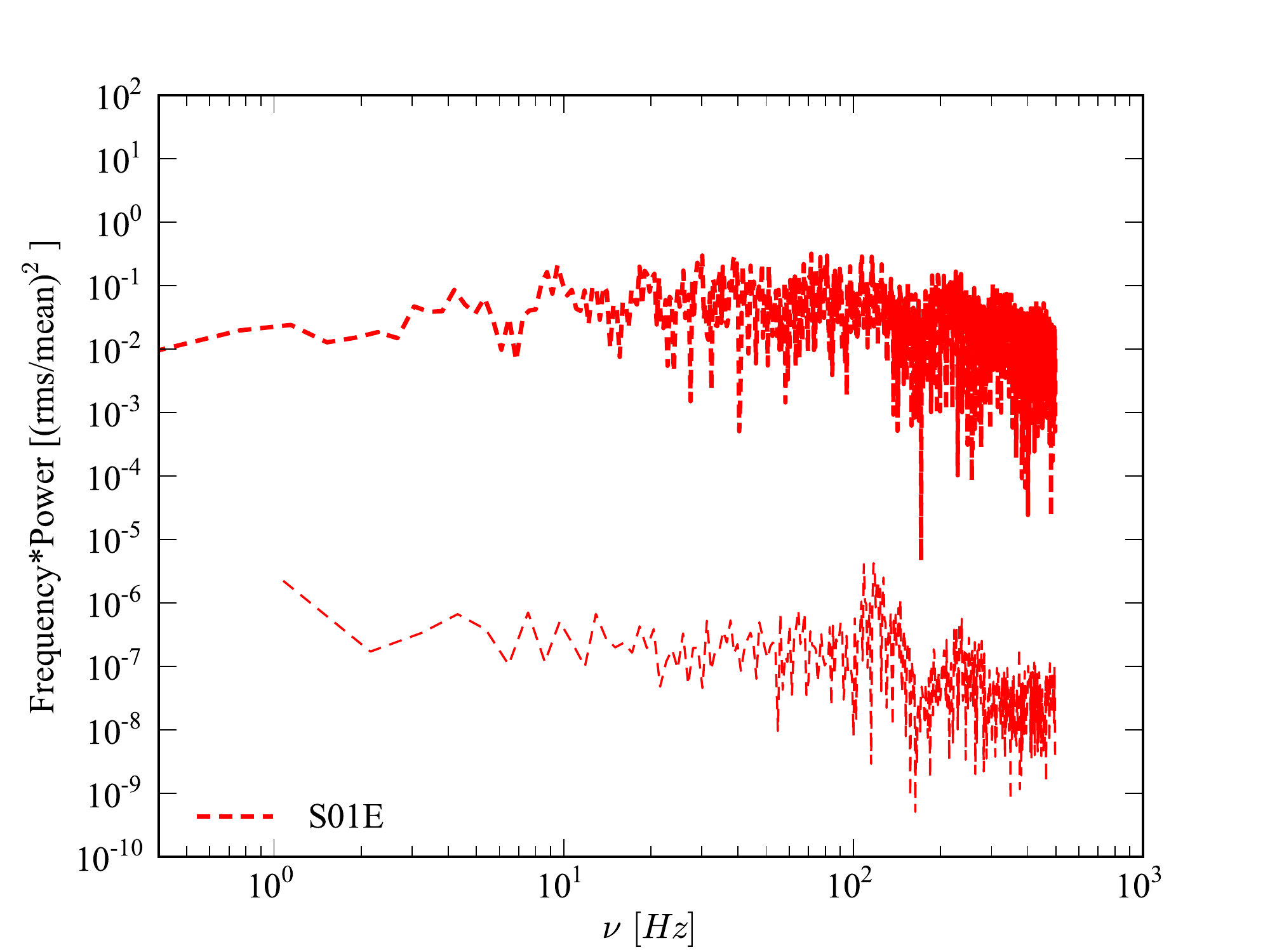} 
\includegraphics[width=0.48\columnwidth]{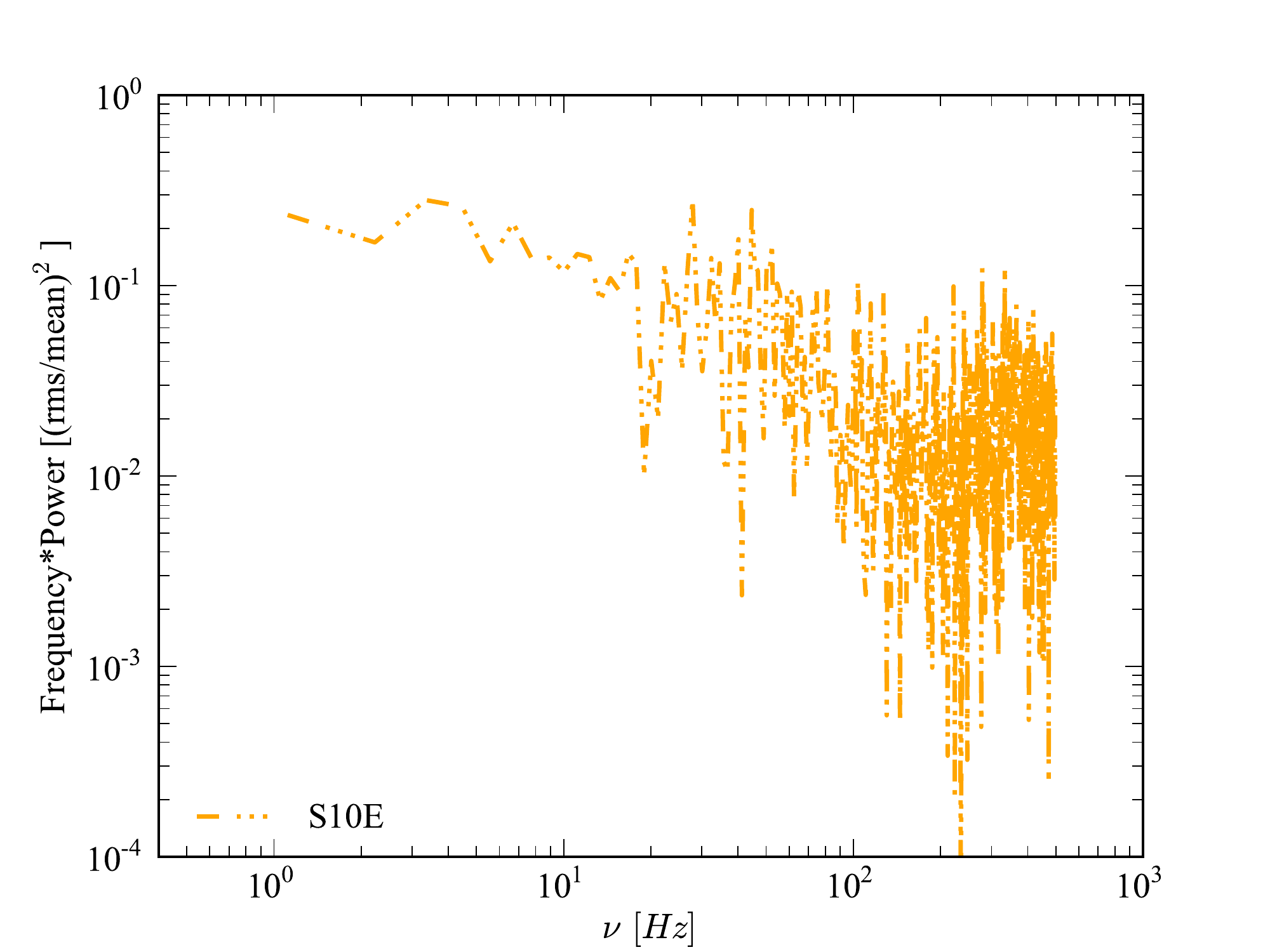} 
\caption{Left: Power spectra corresponding to the unbinned light curve of simulation S01E for the full simulation (thick line) and for the quiescent period from 10000 to $37000\,GM/c^3$ (thin line). Right: Power spectrum for the unbinned light curve of simulation S10E. 
\label{fig:PDS}}
\end{figure}

With light curves in hand, we can now look at their behavior in frequency space to explore the nature of any variability. As we dump data three times per ISCO orbital period (i.e., every 1 ms), we have a Nyquist frequency of 500 Hz. Figure \ref{fig:PDS} gives the Fourier power of the full light curves for simulations S01E and S10E (not the smoothed light curves used for display purposes in Figure \ref{fig:luminosity}). The full power spectra for simulations S01E (thick line, left panel) and S10E (right panel) both appear to have an underlying broken power law dependence on $\nu$, although of differing slopes before and after the break in each case. Interestingly, the break occurs around 100 Hz in both cases. While intriguing, it is not obvious why this frequency, which corresponds to about 3 orbital periods at the ISCO, should be special. None of the power spectra show significant evidence of excess power near certain discrete frequencies, i.e. there is no evidence for quasi-periodic oscillations (QPOs) \citep{Remillard06}.

\begin{figure}
\includegraphics[width=0.9\columnwidth]{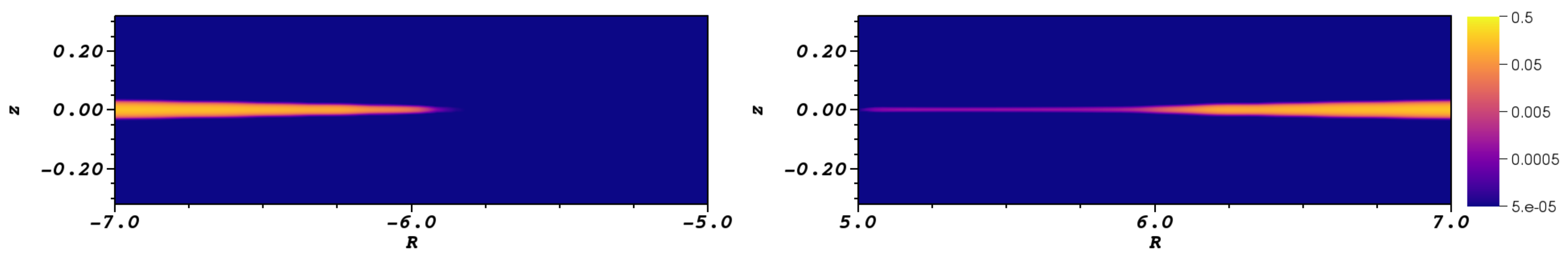} 
\caption{Pseudo-color plots of mass density (cgs units) for simulation S01E. The left- and right-hand panels correspond, respectively, to $t = 38181$ and $t = 38735\,GM/c^3$. While the disk normally truncates close to $6 r_g$ as in the left panel, during the outburst spikes, the disk extends more or less to the inner computational boundary.
\label{fig:isco}}
\end{figure}

Overall, the rms variability in these power spectra are notably high. Remember that in the soft state, BHXRBs often exhibit rms variability $\lesssim 3$\% \citep{vanderKlis04,Done07}. For simulation S10E, the high variability is obvious in the lightcurve in Figure \ref{fig:luminosity}, and is intimately tied to the instability of that disk. However, even the stable, gas-pressure-dominated simulation, S01E, has a surprisingly high fractional rms variability. It seems, though, that most of the power in the S01E spectrum actually comes from trying to fit the luminosity spikes that occur around $t = 3230$, 38700, 51900, 73400, and $74200\,GM/c^3$ in Figure \ref{fig:luminosity}. If, instead, we only analyze the rms variability during a ``quiet'' interval of the S01E lightcurve, say from 10000 to $37000\,GM/c^3$, then we find a much weaker power density spectrum, as shown by the thin, red, dashed line in the left panel of Figure \ref{fig:PDS}. Those spikes in luminosity come from spikes in the mass accretion rate at the inner edge of the disk (note the corresponding spikes in Figure \ref{fig:massFlux}). During these spikes, it seems the disk temporarily fills in the region inside the ISCO. Figure \ref{fig:isco} shows an image of the disk gas density during a normal, quiescent phase (left panel), where the disk is truncated near $6 r_g$, and during one of the accretion spikes (right panel), where the disk extends more or less to the inner edge of the simulation domain ($5 r_g$ in this case). 

The underlying source of true variability in all these simulations seems to be a spectrum of sound waves and diskoseismic modes that are causing fluctuations in the surface density. These modes will be the subject of a future paper.

\section{Discussion \& Conclusions}
\label{sec:conclusion}

In this paper we presented results from a set of numerical simulations of thin-disk accretion onto a Schwarzschild black hole that were designed to be directly comparable to the Shakura-Sunyaev disk model, with particular focus on the unstable, radiation-pressure-dominated branch. As predicted by theory and shown in earlier shearing box simulations, all disks that started on the unstable branch exhibited runaway behavior on timescales comparable to the local thermal time. 

Somewhat surprisingly, though, all of the thermally unstable disks underwent runaway cooling and collapse. We did not see any examples of sustained runaway heating and expansion, even in the case of our simulation S3Ep, for which the disk temperature was intentionally perturbed upward by a factor of 1.5 to try to trigger runaway heating. This tendency of thermally unstable disks to collapse, rather than expand, was also noted earlier in shearing box simulations \citep{Jiang13}. This could indicate that there is something we are still not understanding about this instability. 

We also found that the thermally unstable disks broke up into rings of high and low surface density. Although this behavior is consistent with what is predicted for the viscous instability \citep{Lightman74}, we argued that it is more likely a radiation pressure instability. Whereas the Lightman-Eardley instability predicts an inverse correlation between surface density and vertically integrated stress, we found a positive correlation. Instead, we argued that the low density ``gaps,'' triggered initially by a convective instability, provide preferred channels for radiation to escape, which reinforces the evacuation of these regions. We speculate that in three dimensions, this instability may break the disk up into inhomogeneous clumps of high and low surface density.

Because radiation was included self-consistently in our simulations, we were able to calculate lightcurves and power density spectra corresponding to each simulation. The lightcurves, when coupled with the measured mass accretion rates, revealed radiative efficiencies close to the expected value of 6\% for a Schwarzschild black hole. The rms variability, on the other hand, was seen to be significantly higher in our numerical simulations (up to order unity) than in observed systems ($\lesssim 0.03$). In most cases, this variability was directly related to the instability of the disks. For simulation S01E, the variability was limited to very brief episodes where the disk experienced sharp spikes in mass accretion rate and luminosity. These spikes will be explored further in future work.

Although our results are broadly consistent with theoretical predictions and previous numerical work, they appear to be at odds with observations of BHXRBs. Our unstable simulations -- S1E, S3E, S3Ep, and S10E -- span the exact luminosity range where disks in nature appear to be most stable \citep[e.g.][]{vanderKlis04,Done07}. Furthermore, our one stable simulation -- S01E -- is at a mass accretion rate where BHXRBs appear to be in a spectrally hard state, inconsistent with a simple blackbody disk spectrum \citep{Maccarone03,Remillard06}. In both cases, therefore, it seems there is still important physics that is not being properly captured by standard theory or numerical simulations. Strong magnetic fields have been proposed as a potential solution to the first problem, as such fields may help stabilize radiation-pressure-dominated disks \citep{Begelman07,Oda09,Sadowski16}. Thermal conduction has been suggested as a possible solution to the second problem, by providing a mechanism by which a cold, thin disk may evaporate into a hot, thick flow at low accretion rates \citep{Mayer07}. Each of these will be considered in future work.

\acknowledgements
The authors would like to thank Omer Blaes and Chris Done for useful discussions and feedback regarding this work. This project was supported by National Science Foundation grants AST-1211230, AST-1616185, and PHY-1125915. It used resources from the Extreme Science and Engineering Discovery Environment (XSEDE), which is supported by National Science Foundation grant number ACI-1053575 and PROMETHEUS supercomputer in the PL-Grid infrastructure in Poland.  Additionally, SME acknowledges support from the College of Charleston Undergraduate Research and Creative Activities Board, through SURF grants 2015-10 and 2016-9. Work by PA was performed in part under the auspices of the U.S. Department of Energy by Lawrence Livermore National Laboratory under Contract DE-AC52-07NA27344. BM and WK acknowledge support from Polish NCN grants 2013/08/A/ST9/00795 and 2014/15/N/ST9/04633.

\software{Cosmos++ \citep{Anninos05, Fragile12, Fragile14}}

\end{document}